\documentclass[iop]{emulateapj-rtx4}
\usepackage{amsmath}
\usepackage{graphbox}
\usepackage{bm,xcolor}
\usepackage{threeparttable}
\usepackage{tabularx}
\usepackage{CJK}
\usepackage[colorlinks=true,linkcolor=blue,citecolor=blue]{hyperref}
\def\andname{ }

\def\bfnabla{\ensuremath{\nabla}}

\def\msun{M_\odot}

\def\mbh{M_{\rm{BH}}}

\renewcommand\bv{\boldsymbol{ v}}

\newcommand\bP{\boldsymbol{ P}}
\newcommand\bSr{\boldsymbol{S_r}}
\newcommand\bn{\boldsymbol{ n}}

\newcommand\bI{{\sf\boldsymbol{ I}}}

\newcommand{\kb}{\mathrm{k_B}}

\newcommand\Crat{{\mathbb{C}}}

\def\<{\,\langle\langle}
\def\>{\,\rangle\rangle}

\shorttitle{Prompt Radiation and Mass Outflows From the Stream-Stream Collisions of Tidal Disruption Events}

\begin{document}
\begin{CJK*}{UTF8}{gbsn}

\shortauthors{Y.-F. Jiang et al.}
\author{Yan-Fei Jiang (姜燕飞)\altaffilmark{1,2}, James Guillochon\altaffilmark{1,2}, Abraham Loeb\altaffilmark{1}}

\affil{$^1$Harvard-Smithsonian Center for Astrophysics, 60 Garden Street, Cambridge, MA 02138, USA} 
\altaffiltext{2}{Einstein Fellow}

\title{Prompt Radiation and Mass Outflows From the Stream-Stream Collisions of Tidal Disruption Events}

\begin{abstract}
Stream-stream collisions play an important role for the circularization of highly eccentric streams resulting from tidal disruption events (TDEs). 
We perform three dimensional radiation hydrodynamic simulations to show that stream collisions can contribute significant optical and ultraviolet light 
to the flares produced by TDEs, and can sometimes explain the majority of the observed emission. 
Our simulations focus on the region near the radiation pressure dominated shock produced by a collision and track how the kinetic energy of the stream is dissipated by the associated shock. 
When the mass flow rate of the stream $\dot{M}$ is a significant fraction of the Eddington accretion rate, $\gtrsim2\%$ of the initial kinetic energy is converted to radiation directly as a result of the collision. 
In this regime, the collision redistributes the specific kinetic energy into the downstream gas and more than $16\%$ of the mass can become unbound. 
The fraction of unbound gas decreases rapidly as $\dot{M}$ drops significantly below the Eddington limit, with no unbound gas being produced when 
$\dot{M}$ drops to $1\%$ of Eddington; we find however that the radiative efficiency increases slightly to $\lesssim 8\%$ 
in these low $\dot{M}$ cases. The effective radiation temperature and size of the photosphere is determined by the stream velocity and $\dot{M}$, 
which we find to be a few times $10^4$~K and $10^{14}$~cm in our calculations, comparable to the inferred values of some TDE candidates. 
The photosphere size is directly proportional to \smash{$\dot{M}$}, which can explain the rapidly changing photosphere sizes seen in TDE candidates such as PS1-10jh.
\end{abstract}

\maketitle

\section{Introduction}
A star can be tidally disrupted by a galaxy's central supermassive black hole (SMBH) when it passes 
within the tidal radius $r_{\rm t} \equiv R_{\ast} (M_{\rm h} / M_{\ast})^{1/3}$, where $R_{\ast}$ and $M_{\ast}$ are the star's radius and mass, and $M_{\rm h}$ is the mass of the black hole. The candidates of tidal disruption events (TDEs) are 
usually found by searching for flares co-located with the galactic nucleus in the X-ray, UV and
optical bands \citep[][and references therein]{KomossaBade1999,Donleyetal2002,Gezarietal2006,Gezarietal2009,Bloom:2011a,
VanVelzenetal2011,Gezarietal2012,Arcavietal2014,Vinko:2015a,Komossa2015,Lin:2015a,Holoien:2015a}. The flares 
typically reach a peak luminosity $\sim 10^{43}-10^{44}$ erg s$^{-1}$ within 
$\sim$ months, which corresponds to the Eddington luminosity of a 
$\sim 10^5-10^6\msun$ black hole. 

The flares of TDEs have been thought to be produced from accretion disks that form
after the stellar debris circularizes \citep[][]{Rees1988,StrubbeQuataert2009}. For the stellar stream 
with a flat energy distribution per unit mass, the mass-return rate decreases with time 
as $\propto t^{-5/3}$ \citep[][]{Phinney1989}. It is usually assumed that the stellar stream 
can be circularized and accreted quickly so that the observed luminosity also declines with 
time as $\propto t^{-5/3}$, and this has been adopted as a hallmark of many observed 
TDEs. However, the effective thermal temperature of the observed flares is only a few$\times 10^4$ K 
and remains approximately constant when the luminosity declines \citep[][]{Gezarietal2012,Arcavietal2014}. 
This is difficult to explain if the flares are indeed produced from the standard accretion disks, whose temperatures should vary with varying rates of accretion \citep{Beloborodov:1999a}.

Significant progress has been made recently in studying the detailed processes in TDEs based on 
grid based or smooth particle hydrodynamic simulations \citep[][]{Rosswog:2009a,Guillochonetal2013,
Guillochon:2014b, Shiokawa:2015a,
Hayasakietal2015,Sadowski:2015b,Bonnerotetal2016}. These studies raise more challenges in the classical picture of 
TDEs. The simulations usually find that it is very difficult to circularize the stellar streams if the orbits remain parabolic, as might be expected when the orbital dynamics are Newtonian. 
Additionally, the distribution of mass per unit energy is not exactly flat \citep[][]{Lodato:2009a,Guillochonetal2013}, and when observed at a single frequency (such as a photometric band), the light produced by an accretion disk resulting from a TDE may not follow the bolometric luminosity \citep{Lodato:2011a}. All these effects will make the radiation emitted by any formed accretion disks deviate significantly from the $t^{-5/3}$ law, in contrast to the observed lightcurves of TDE candidates which appear to follow the fallback rate, even when observed in a single band. 

Alternative mechanisms for describing the appearance of TDEs that can solve some of these puzzles include a hydrostatic envelope around the black hole emitting at the
Eddington limit \citep[][]{Loeb:1997a,CoughlinBegelman2014}; possibly in conjunction with powerful disk winds that limit the accretion rate 
\citep[][]{StrubbeQuataert2009,Metzger:2015a,Miller2015}.  \cite{Piranetal2015} and \cite{Svirski:2015a} have proposed 
that the shock near the apocenter can also produce the observed TDE emission. However, there have been 
no quantitive calculations showing how the photons are produced by this mechanism and whether the emission is consistent with the observed 
properties of TDEs. 
As shocks caused by the stream-stream collisions 
clearly play an important role to convert the kinetic energy of the streams into thermal energy \citep[][]{Rees1988, 
Kochanek:1994a}, we wish to study here in detail the structure of the shock when the stream first collides with itself 
and how this conversion proceeds. We also want to evaluate whether the stream-stream collision will naturally produce 
an emitting structure with a size on the order of hundreds of AU and UV temperatures as has been observed for the thermal TDE candidates.

For the stellar streams on a parabolic orbit around the black hole, the kinetic energy is comparable to the 
gravitational potential energy. When a non-negligible fraction of the kinetic energy is converted to the thermal 
energy, the downstream gas is likely to be radiation pressure dominated \citep[][]{Kimetal1999}. 
All previous studies are based on hydrodynamic simulations with an assumed equation of state, and the thermal properties of the stream and the effects of strong radiation pressure are all neglected. In fact, 
the shocks are likely unresolved in most previous simulations. As \cite{Hayasakietal2015} noted, 
the thermal properties of streams affect the circularization process significantly, and in this paper we consider streams that are initially ``cold,'' in contrast to the wide fans that are only realized at low mass ratios \citep[e.g.][]{Rosswog:2009a,Guillochon:2014a,Shiokawa:2015a} or deeply-penetrating encounters \citep[e.g.][]{Sadowski:2015b}. 
By using an accurate radiative transfer algorithm \citep{Jiangetal2014b}, we can study here the thermal properties of the 
radiation-pressure-dominated shock self-consistently for the first time. 

The paper is organized as follows. The typical properties of the colliding streams resulting from tidal disruptions and how they collide with one another is 
described in Section \ref{sec:whereandhow}. We describe the equations we solve and the numerical setup we use 
in Section \ref{sec:method}. Our quantitative results, including radiative efficiencies and outflow rates, are provided in Section \ref{sec:results}. In Section 
\ref{sec:discussion} we discuss the implications of our calculations and some directions for future study. 

\section{Where and how streams collide}
\label{sec:whereandhow}

When a star is disrupted by a SMBH, the debris stretches away from the black hole in a highly-elongated stream. Despite recent assertions to the contrary \citep{Krolik:2016a}, cases where the star is not shock-heated during its closest approach at a pericenter distance $r_{\rm p}$ \citep[ with $\beta\equiv r_{\rm t}/r_{\rm p} \lesssim 3$, which constitutes 80\% of tidal disruptions,][]{Carter:1985a,Guillochon:2009a} result in post-disruption debris that is self-gravitating, with a width $h$ that only grows modestly with distance from the black hole, $h \propto r^{1/4}$ \citep{Kochanek:1994a,Guillochon:2014a,Coughlin:2015a,Coughlin:2016a}. In the simulation of \citet{Shiokawa:2015a}, the self-gravity of the matter is explicitly switched off shortly after disruption, resulting in a debris stream that quickly broadens due to the stream's internal pressure gradient. This stream is artificially wider (by orders of magnitude) than it should be at the moment of self-interaction than the more-realistic case where self-gravity is included, with the end result resembling much rarer ``deep'' encounters in which the stream is broadened by the injection of heat at periapse. Even in cases where the stream passes through periapse multiple times before colliding due to out-of-plane precession effects, the streams remain much thinner than in simulations where self-gravity is ignored \citep{Hayasakietal2015}. Because collisions between two thin streams are more likely to occur given the smaller rate of deep events, we ignore the less common fanned-stream scenarios. However, given that these deep events are likely to be associated with pre-disruption X-ray flashes \citep{Guillochon:2009a} and strong gravity environments that can produce detectable gravitational waves \citep{Kobayashi:2004a}, they deserve a full radiation-transport treatment as well, which we leave to future work.

For non-spinning black holes, the location of the first stream-stream intersection point can be determined \citep{Dai:2015a} from the eccentricity of the most-bound debris
\begin{equation}
e_{\rm mb} = 1 - 0.02 (M_{\ast}/M_{\odot})^{1/3} (M_{\rm h}/10^{6} M_{\odot})^{-1/3} \beta^{-1}
\end{equation}
and the amount of instantaneous de Sitter precession at periapse $\Omega$ \citep{Guillochon:2015b}
\begin{equation}
\Omega = \frac{6 \pi G M_{\rm h} \beta}{c^{2} q^{1/3} R_{\ast} (1 + e_{\rm mb})},\label{eq:omega}
\end{equation}
where $q \equiv M_{\odot} / M_{\ast}$, with the distance of collision
\begin{equation}
r_{\rm c} = \frac{\left(1 + e_{\rm mb}\right) r_{\rm t}}{\beta \left[1 - e_{\rm mb} \cos \left(\Omega/2\right) \right]}.
\end{equation}
Assuming $e_{\rm mb} \simeq 1$ and $\Omega \ll \pi$, the expression for $r_{\rm c}$ simplifies to
\begin{equation}
r_{\rm c} = \frac{16}{\Omega^{2}} \frac{r_{\rm t}}{\beta}.
\end{equation}
Substituting the definition for $\Omega$ from Equation (\ref{eq:omega}) and scaling to fiducial values, we find
\begin{equation}
\frac{r_{\rm c}}{r_{\rm s}} \simeq 12 \beta^{-3} \left(\frac{R_{\ast}}{0.5 R_{\odot}}\right)^{3} \left(\frac{M_{\rm h}}{10^{7} M_{\odot}}\right)^{-2},\label{eq:rc}
\end{equation}
where we have taken the fiducial stellar radius $R_{\ast}$ to be equal to that of a $0.3 M_{\odot}$ star, approximately the median mass of the standard Kroupa IMF \citep{Kroupa:2001a}. As the streams move on nearly-parabolic orbits, the speed of the fluid within the stream at the collision point is
\begin{equation}
\frac{v_{\rm c}}{c} = 0.3 \beta^{3/2} \left(\frac{R_{\ast}}{0.5 R_{\odot}}\right)^{-3/2} \left(\frac{M_{\rm h}}{10^{7} M_{\odot}}\right).
\end{equation}
The angle of collision is given by \citep{Dai:2015a}
\begin{equation}
\theta_{\rm c} = \frac{1 - 2 \cos (\Omega/2) e_{\rm mb} + \cos (\Omega) e_{\rm mb}^{2}}{1 - 2 \cos (\Omega/2) e_{\rm mb} + e_{\rm mb}^{2}},
\end{equation}
which simplifies again assuming $e_{\rm mb} \simeq 1$ and $\Omega \ll \pi$ to
\begin{equation}
\theta_{\rm c} \simeq \pi - \beta \left(\frac{M_{\rm h}}{10^{7} M_{\odot}}\right)^{2/3} \left(\frac{R_{\ast}}{0.5 R_{\odot}}\right),\label{eq:thetac}
\end{equation}
which $\simeq \pi - 1 = 123^{\circ}$ for the fiducial values. $\theta_{\rm c}$ is typically larger than $90^{\circ}$ except for the most relativistic encounters, meaning that the two streams interact violently at their point of intersection, converting a large fraction of the kinetic energies of both streams into heat.

\subsection{Vertically offset stream collisions}\label{sec:offset}
For spinning black holes, there is no apriori reason to expect the incoming star's orbital angular momentum vector to align with the black hole's spin vector, and thus the streams are likely to experience some degree of out-of-plane precession resulting from frame-dragging effects \citep{Stone:2012a}. The magnitude of this out-of-plane precession is
\begin{equation}
\Phi = \frac{4 \pi a  \sin i}{c^{3} q^{1/2}} \left[\frac{G M_{\rm h} \beta}{R_{\ast} (1 + e_{\rm mb})}\right]^{3/2}
\end{equation}
where $c$ is the speed of light, $a$ is the black hole's dimensionless spin parameter, and $i$ is the inclination of the star's orbit to the black hole's spin plane. Once the debris stream crosses periapse (the second crossing of the matter by periapse after the initial disruption), the effects of self-gravity within the stream are minuscule and the stream's diameter is instead dictated by its internal pressure and the black hole's tidal gravity. If the stream's internal temperature remains low, as is expected for the highly-eccentric disruptions that likely dominate the tidal disruption rate \citep{Hayasakietal2015}, the stream's diameter when it returns to periapse can be even thinner than the star's original diameter. Multiplying this by $r_{\rm c}$ in Equation (\ref{eq:rc}), we find a typical collision offset of
\begin{equation}
\frac{z_{\rm c}}{H(r_{\rm c})} = 60 a \beta^{3/2} \left(\frac{M_{\rm h}}{10^{7} M_{\odot}}\right)^{4/3} \left(\frac{R_{\ast}}{0.5 R_{\odot}}\right)^{-3/2} \sin i \label{eq:offset}
\end{equation}
where $H(r_{\rm c}) = R_{\ast} (r_{\rm c} / r_{\rm p})$ is the stream's scale-height at the distance of the intersection $r_{\rm c}$. For deep encounters of small stars about large, maximally-spinning black holes, it is clear that the streams will frequently miss each other as the deflection angle is significantly larger than the size of the streams, in which case the onset of circularization may be significantly delayed, resulting in a ``dark year'' \citep{Guillochon:2015b}. When the streams eventually do collide after the delay, they are more likely to do so close to the black hole at $r \sim r_{\rm p}$ where they experience most of their precession. For grazing encounters of larger stars about lower-mass, lower-spin black holes,
\begin{align}
\frac{z_{\rm c}}{H(r_{\rm c})} &= 0.1  \left(\frac{a}{0.1}\right) \left(\frac{\beta}{0.5}\right)^{3/2}\nonumber\\
&\times \left(\frac{M_{\rm h}}{10^{6} M_{\odot}}\right)^{4/3} \left(\frac{R_{\ast}}{R_{\odot}}\right)^{-3/2} \sin i,
\end{align}
and thus the offset can be much less than $H$, resulting in a nearly direct stream intersection.

The variety of possible angles, offsets, and collision locations as the black hole mass, stellar mass, and stellar size change suggests that the circularization process likely differs significantly from event to event. This motivates running separate simulations where these parameters are varied over a reasonable range. In this paper, we hold the location of the collision point fixed, and vary other parameters of the stream self-intersection, namely the angle of the collision, whether the stream collision is head-on or at some vertical offset as well as a range of mass flow rates. Our numerical setup is described below.

\begin{figure*}
\begin{center}
\includegraphics[width=1.0\linewidth]{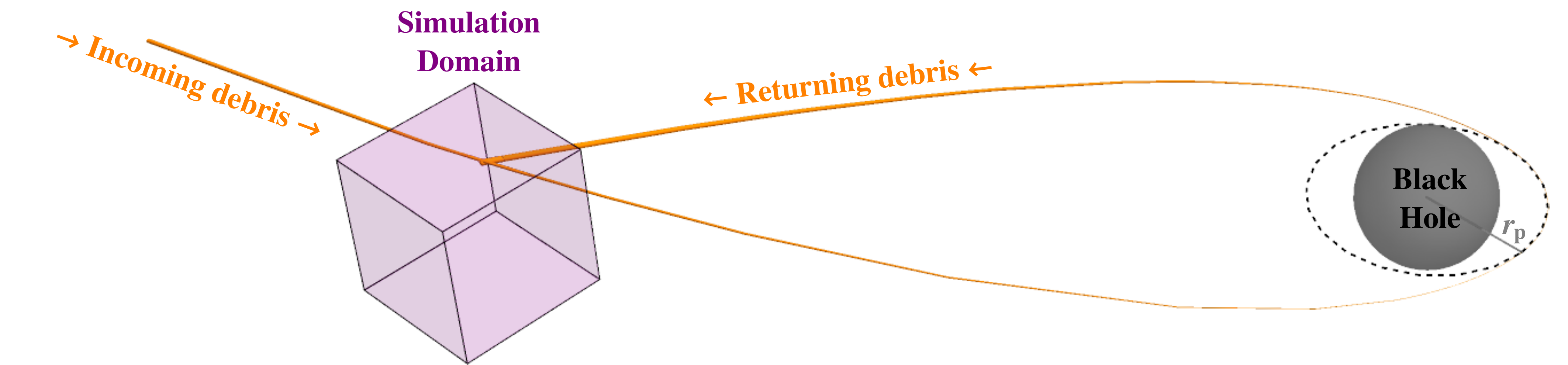}
\caption{An example setup of our simulation domain with respect to the global geometry of the streams. The parameter combination shown here ($M_{\rm h} = 10^{6.95} M_{\odot}$, $M_{\ast} = 0.18 M_{\odot}$, $\beta = 0.64$, $a = 0.035$) is drawn from the Monte Carlo results of \citet{Guillochon:2015b}, and results in an angle of stream self-intersection of 138$^{\circ}$.}
\label{setup}
\end{center}
\end{figure*}

\section{Numerical Method}
\label{sec:method}
\subsection{Equations}

We solve the frequency-independent (gray)
radiation hydrodynamic equations in cartesian coordinates $(x,y,z)$
with unit vectors ($\boldsymbol{ \hat{x}}$, $\boldsymbol{\hat{y}}$, $\boldsymbol{\hat{z}}$). 
Our calculations employ the new radiation MHD code {\sc Athena++} (Stone et al., 2016, in preparation), 
which is an extension of the multidimensional MHD code 
{\sc Athena} \citep[][]{Stoneetal2008}.
The gravitational acceleration due to the central black hole is neglected, 
as we only focus on the collision point of the stream and the 
simulation domain is significantly smaller than the distance of the collision point to the black hole (see Figure~\ref{setup}). 
With the radiation momentum and energy source terms $\bSr\left(\bP\right), S_r(E)$, 
the equations we solve are \citep[][]{Jiangetal2014b}
\begin{eqnarray}
\frac{\partial\rho}{\partial t}+\bfnabla\cdot(\rho \bv)&=&0, \nonumber \\
\frac{\partial( \rho\bv)}{\partial t}+\bfnabla\cdot({\rho \bv\bv+P\bI}) &=&-\bSr(\bP),\  \nonumber \\
\frac{\partial{E}}{\partial t}+\bfnabla\cdot\left[(E+P)\bv)\right]&=&-cS_r(E),  \nonumber \\
 \frac{\partial I}{\partial t}+c\bn\cdot\bfnabla I&=&S.
\label{Equation}
\end{eqnarray}
In the above equations, $\rho$, $P$, $\bv$, $\bI$ are the gas density, pressure,
flow velocity, and the unit tensor.  
The total gas energy density is $E=E_g+\rho v^2/2$, where $E_g=P/(\gamma-1)$ is the
internal gas energy density with a constant adiabatic index
$\gamma=5/3$.  The gas pressure is $P=\rho \kb T/\mu$, where
$\kb$ is Boltzmann's constant, $T$ is the gas temperature, $\mu=0.62m_p$
is the mean molecular weight for fully ionized gas of solar metallicity and 
$m_p$ is the proton mass. 

The radiation source term $S$ in the co-moving frame of the fluid is 
\begin{eqnarray}
S=c\rho\kappa_a\left(\frac{a_rT^4}{4\pi}-I_0\right)+c\rho\kappa_s\left(J_0-I_0\right), 
\end{eqnarray}
where $\kappa_a$ and $\kappa_s$ are the absorption and scattering opacities, 
while $J_0=\int I_0d\Omega$ is the angular quadrature of the specific intensity 
in the co-moving frame. 
The radiative transfer equation is solved with a similar method as described in \cite{Jiangetal2014b}, 
which has been successfully used to study the super-Eddington black hole accretion 
disks \citep[][]{Jiangetal2014c}, where the algorithm has been extended to include all $v/c$ terms 
in {\sf Athena++}. 
The lab frame specific intensity $I$ is first transformed to the co-moving frame $I_0$ via Lorentz transformation 
and $I_0$ is updated implicitly with the source terms in the co-moving frame. 
The updated $I_0$ is then transformed back to the lab frame. All the velocity-dependent 
terms are handled automatically via Lorentz transformations and the other parts of the algorithm 
are the same as in \cite{Jiangetal2014b}. 
In the calculations presented in this work, we only include the electron scattering opacity, $\kappa_{\rm es}=0.34$ cm$^2$/g, 
and free-free absorption opacity, 
$\kappa_a=2.86\times 10^{-5}\left[\rho/\left(10^{-8} {\rm g/cm^{3}}\right)\right]\left[T/\left(10^6{\rm K}\right)\right]^{-3.5}$ 
cm$^2$/g,
 as in \cite{Jiangetal2014c}.  Effects of additional opacities caused by the metals, which are clearly important 
for the observational appearance of TDEs \citep[][]{Roth:2015a,Kochanek:2015a} 
and the formed accretion disks \citep[][]{Jiangetal2016}, will be studied in the future. 

\subsection{Simulation Setup}
\label{sec:setup}
We inject two streams into the simulation box through the left $x$ and $y$ boundaries, which 
represent the self-collision part of the same stream (Figure~\ref{setup}), where the stream is assumed to be formed from a disruption of a low-mass star by a supermassive black hole. Because the collision can occur at a variety of distances from the black hole at various velocities and angles, the setup we have chosen can simultaneously apply to very different physical systems, as many will coincidentally share the same conditions at the stream collision point. Figure~\ref{setup} shows the size of the simulation domain relative to the disruption of a $M_{\ast} = 0.2 M_{\odot}$ star that is tidally 
disrupted by $\mbh=10^{7}\msun$ black hole with a periapse distance $r_{\rm p}\simeq 2r_{\rm s}$, where $r_{\rm s}=2.63\times 10^{12}$ cm 
is the Schwarzschild radius of the black hole, for which the collision distance $r_{\rm c}=10r_{\rm s}$, which gives the 
incoming stream velocity $v_i=\sqrt{2G\mbh/r_{\rm c}}=0.31 c$ for a stream initially on a parabolic orbit.

The typical scale height 
of the stream $H$ is determined by the perpendicular component of the tidal force of the black hole at $r_{\rm c}$ and the gas pressure 
gradient \citep[][]{Guillochon:2014a}
\begin{eqnarray}
H\approx 2r_{\rm c}\sqrt{\frac{r_{\rm c}}{r_{\rm s}}\frac{\kb T_{\rm i}}{\mu c^2}}=7.37\times 10^{9}\ \sqrt{\frac{T_{\rm i}}{10^6 {\rm K}}}\ {\rm cm},
\end{eqnarray}
where $T_{\rm i}$ is the initial temperature of the stream before the collision. Hydrodynamic simulations of TDEs \citep[][]{Hayasakietal2015} 
suggest that $T_{\rm i}\sim 10^4-10^6$ K at the time of collision. Because the kinetic energy of the stream is $3\times 10^6$ times the internal energy for $T_{\rm i}=10^5$ K, 
the exacted initial temperature has negligible effects on the properties of the stream in the downstream of the shock after the collision. As there is no 
gravity in our simulations, $H$ is a free parameter, and for 
numerical reasons we choose $T_{\rm i}=10^6$ K to set the thickness $H$ so that the incoming stream is well-resolved.
The gas temperature of incoming stream is chosen to $10^5$ K so that it is much colder than the downstream gas. 
The density profile of  the stream is taken to be $\rho=\rho_0\exp(-0.5s^2/H^2)$, where $s$ is the distance to the center of the stream and $\rho_0$ is the density 
at the center. We assume a velocity profile across the two streams that is mostly constant with $v = v_{\rm i}$ but with a sharp decline at a distance $\gtrsim 2 H$ from the stream center, $v=v_i\exp[-\left(s/2H\right)^{16}]$, this sharp cutoff is chosen to avoid numerical issues in the low density region immediately exterior to the streams.
The mass flow rate with each stream is $\dot{M}/\dot{M}_{\rm Edd}\approx 2\pi H^2\rho_0 v_i/\dot{M}_{\rm Edd}=120\rho_0r_{\rm s}\kappa_{\rm es}/\Crat^2$, 
where $\kappa_{\rm es}$ is the electron scattering opacity, $\dot{M}_{\rm Edd}=40\pi G\mbh/\left(c\kappa_{\rm es}\right)$ is the Eddington accretion rate 
with an assumed $\epsilon = 10\%$ efficiency, and $\Crat\equiv c/\sqrt{\kb T_{\rm i}/\mu}=2.55\times 10^3/\sqrt{T_{\rm i}/10^6 {\rm K}}$ is the ratio between the speed of light and the isothermal 
sound speed at a temperature $T_{\rm i}$. The actual mass flow rate of streams from TDEs varies in a large range for totally and partially disrupted stars of different masses and stellar types \citep[][]{MacLeod:2012a,Guillochonetal2013}, and varies in time for a given even by many orders of magnitude. A typical value for the density at the stream center given an Eddington ratio $\dot{M}/\dot{M}_{\rm Edd} \sim 0.1$ for the example shown in Figure~\ref{setup} is $\rho_0 \sim 10^{-8}$ g cm$^{-3}$, which is very optically thick ($r_{\rm s}\rho_0\kappa_{\rm es} \sim 10^4$) as a result of the mass being funneled through a narrow stream. In our simulations, we vary the mass flow rate to determine how our results change with changing fallback rates. 

The streams will collide with angles usually in the range $\theta\sim 40^{\circ}-160^{\circ}$ due to relativistic precession \citep[][]{Kochanek:1994a,Kimetal1999,
Guillochon:2015b,Dai:2015a}. We 
choose two examples with $\theta=135^{\circ}$ and $90^{\circ}$ to show the effects of varying the collision angle, which are labeled as {\sf A1} and {\sf A2}.
For a non-spinning black hole, the streams will collide in the orbital plane, and we have made this assumption in cases {\sf A1} 
and {\sf A2}. With black hole spin, the precession is no longer confined to the orbital plane, which can cause a vertical offset when the streams collide (see Section~\ref{sec:whereandhow}). To treat 
this case, we apply a vertical offset $z_{\rm c} = H$ to simulation identical to {\sf A1} while keep all the other parameters fixed, which is labeled as {\sf A1z}. 
To study the cases with different injected mass flow rates, runs {\sf A1m1}, {\sf A1m2} and {\sf A1m3} have $200\%$, $40\%$ and 
$10\%$ mass flow rates as in {\sf A1}, while the other parameters are the same as in {\sf A1}. 
The box sizes of all the simulations are fixed to be $L_x=L_y=L_z=2r_{\rm s}$ with resolution $N_x=N_y=N_z=384$, and thus the two injected streams are initially resolved with $10$ cells 
across each of their diameters; once the streams collide with one another and expand downstream, the post-collision region is well-resolved.
The characteristic timescale is the stream crossing time $t_0\equiv r_{\rm s}/v_i=3.19\times 10^2$ s, which we will use as our fiducial time unit. 
The fiducial density unit is $\rho_0=5.94\times 10^{-9}$ g/cm$^3$ for all the simulations. 
We use $80$ angles per cell for specific intensity $I$ 
to resolve the angular distribution of the radiation field. The streams are injected at $x=-r_{\rm s}, y=-0.72r_{\rm s}, z=0$ and $x=-0.72r_{\rm s}, y=-r_{\rm s}, z=0$ for {\sf A1}, 
 $x=-r_{\rm s}, y=-0.8r_{\rm s}, z=0$ and $x=-0.8r_{\rm s}, y=-r_{\rm s}, z=0$ for {\sf A2}.
For {\sf A1z}, the streams are injected at $x=-0.72r_{\rm s}, y=-r_{\rm s}, z=-0.5H$ and $x=-r_{\rm s}, y=-0.72r_{\rm s},z=0.5H$. Within $2H$ of the injection region, the ghost zones 
are fixed to be the initial condition of the incoming streams. In all the other regions of the six faces, we copy all the variables in the last active zones to the ghost 
zones, and we do not allow inflow except within the stream injection regions.



\section{Results}
\label{sec:results}

\subsection{Simulation History}

\begin{figure}[t]
\begin{center}
\includegraphics[width=1.0\columnwidth]{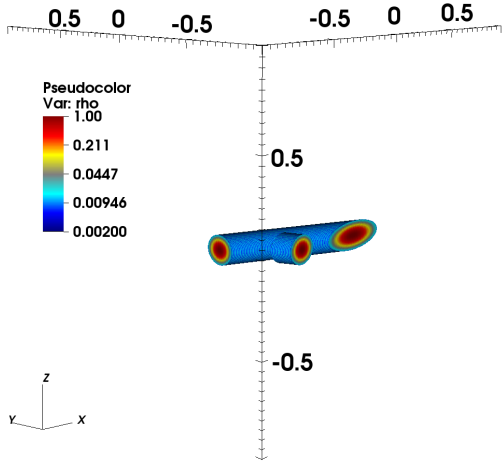}
\caption{Density of the initial streams before the collision for run {\sf A1}. 
The streams are injected from left $x$ and $y$ boundaries as described in Section \ref{sec:setup}, which 
represents the self intersection part of the same stream from TDEs. Unit of the box size is $r_{\rm s}$ while 
the density unit is $\rho_0=5.94\times 10^{-9}$ g/cm$^3$.}
\label{Inistream}
\end{center}
\end{figure}

The initial streams we inject from the boundaries as described in Section \ref{sec:setup}  for run {\sf A1} 
are shown in Figure~\ref{Inistream}. For all runs, the stream resulting from the disruption is injected 
from the left $x$ boundary and is presumed to leave the domain from the right $x$ boundary, returning to the domain through the left $y$ boundary. The returning stream then intersects with 
itself at an angle $\theta$, which is $135^{\circ}$ for {\sf A1}. The streams collide and a strong radiation 
pressure dominated shock is formed, which converts the steam's kinetic energy to thermal energy. 
The detailed structures formed in the post-shock region will be studied in next section.
Due to momentum conservation, the mean motion of the downstream gas is along the direction 
of the net velocity of the original two streams, which have the same density and velocity magnitude. 
At the same time, the strong downstream 
radiation pressure accelerates the gas and causes the gas to expand. Most of the thermal 
energy is converted back to the kinetic energy within this expanding plume. The remaining thermal energy 
is emitted near the photosphere of the plume where it can contribute to the light emitted by the TDE.

\begin{figure}[t]
\begin{center}
\includegraphics[width=1.0\columnwidth]{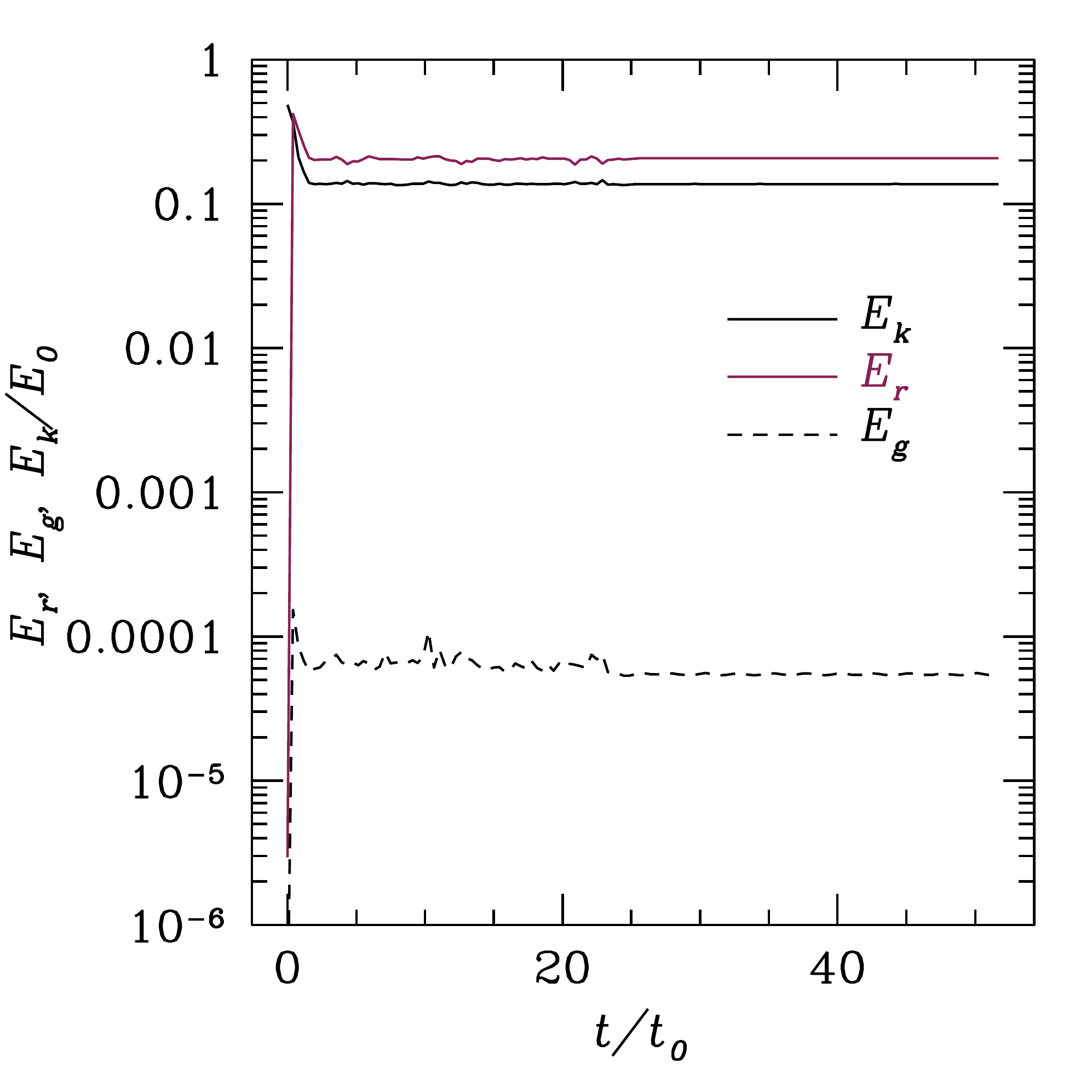}
\caption{Volume-averaged, density-weighted kinetic energy density $E_k$ (solid black line), 
radiation energy density $E_r$ (red line) and gas internal energy $E_g$ (dashed black line) 
as a function of time $t$ for run {\sf A1}. The unit used for the energy densities is $E_0=\rho_0v_i^2/2$ 
while the time unit is $t_0=r_{\rm s}/v_i=3.2\times 10^2$ s.}
\label{EnergyHistory}
\end{center}
\end{figure}

Histories of the density-weighted averaged kinetic energy density $E_k$, radiation energy density $E_r$ 
and gas internal energy $E_g$ for {\sf A1} are shown in Figure~\ref{EnergyHistory}. 
The thermal 
energies ($E_r$ and $E_g$) initially rapidly increases while the kinetic energy drops due to the shock 
within a single stream sound crossing time $t_0$, with the radiation pressure being three orders of magnitude larger than the gas pressure. 
Afterwards, the system settles down to a steady state within a few $t_0$. The injected energy flux by the incoming 
stream is balanced by the outgoing energy fluxes carried by the downstream gas, with the density-weighted energy densities being dominated by the downstream region of the shock. 
The fact that $E_r$ is comparable to $E_k$ means 
the kinetic energy is efficiently converted to the thermal energy in this case. 
The fluctuations in the steady state are caused by the unsteady shock front, which will be discussed in the next section. 

\subsection{Stream Structures after the Shock}
\label{sec:shock}

\begin{figure}[t]
\begin{center}
\includegraphics[width=1.0\columnwidth]{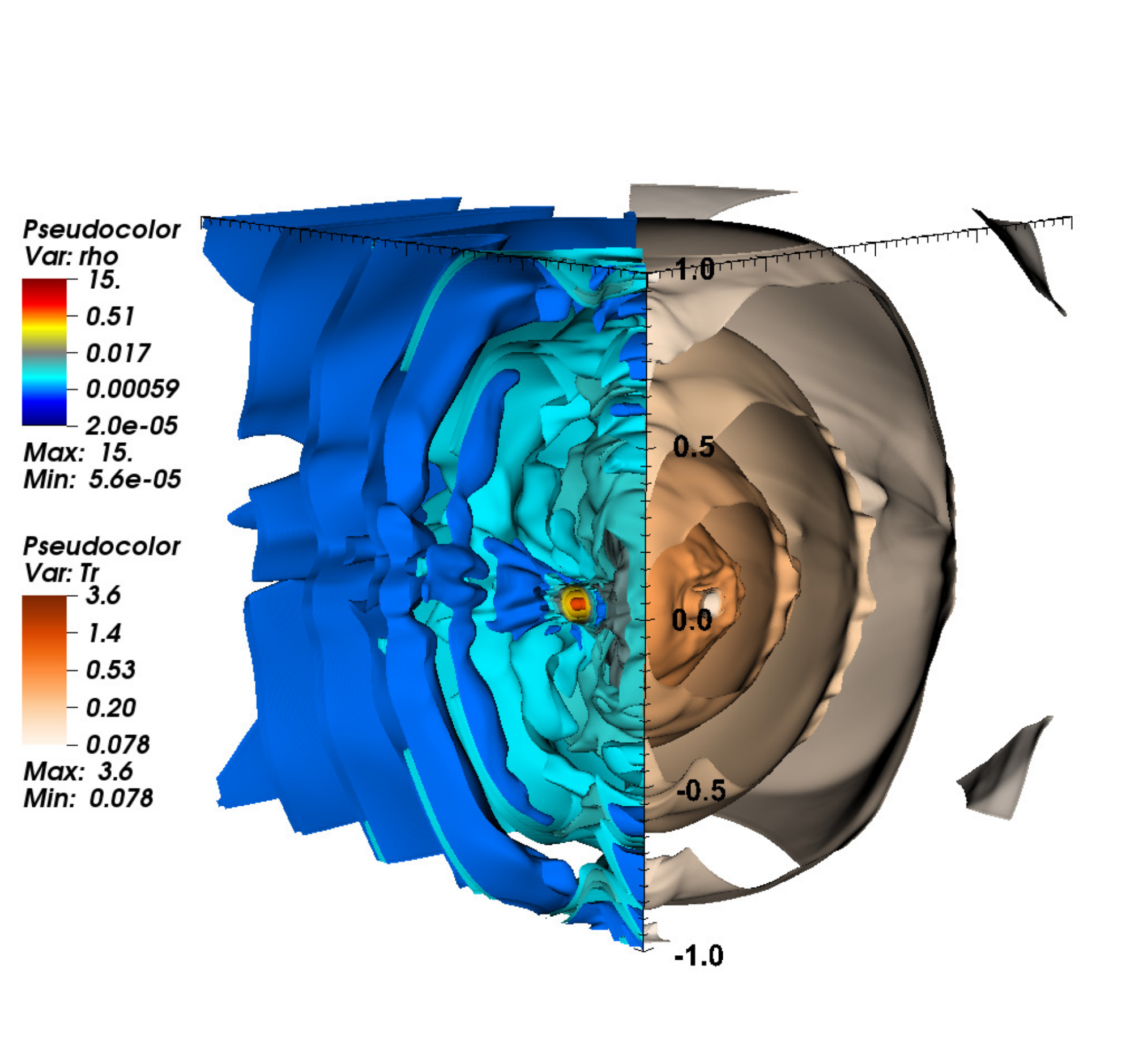}
\caption{Snapshot of the 3D density $\rho$ (left) and radiation temperature $T_{\rm r}$ (right) 
for run {\sf A1} at time $29.6t_0$. Density unit is $\rho_0=5.94\times 10^{-9}$ g/cm$^3$ 
while the temperature unit is $T_{\rm i}=10^6$ K. The length unit is $r_{\rm s}$. 
(A movie showing the density evolution of this run is available at \url{
https://youtu.be/vc7O10UVFxg}.)}
\label{A1RhoTr}
\end{center}
\end{figure}

\begin{figure}[t]
\begin{center}
\includegraphics[width=1.0\columnwidth]{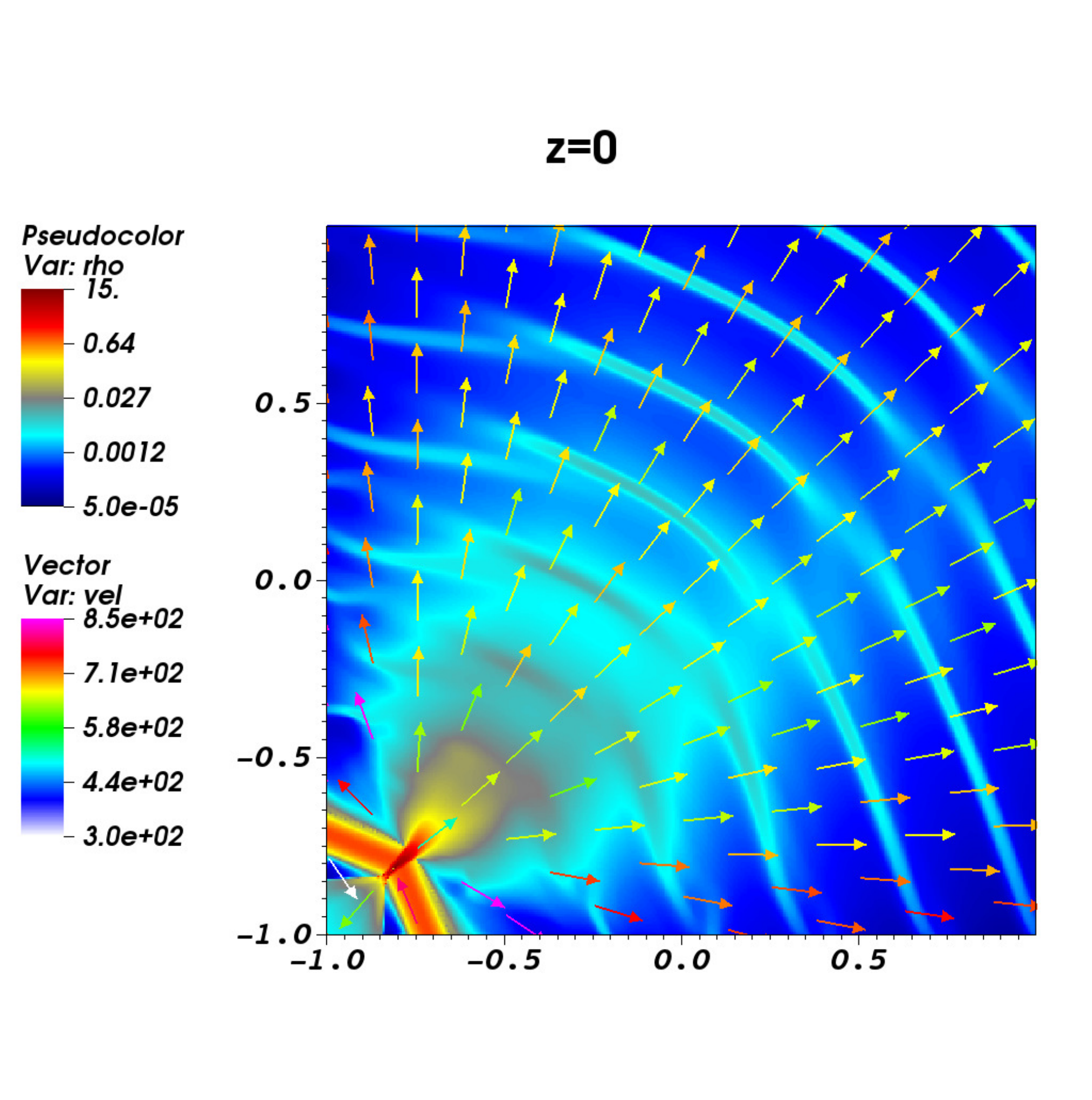}
\includegraphics[width=1.0\columnwidth]{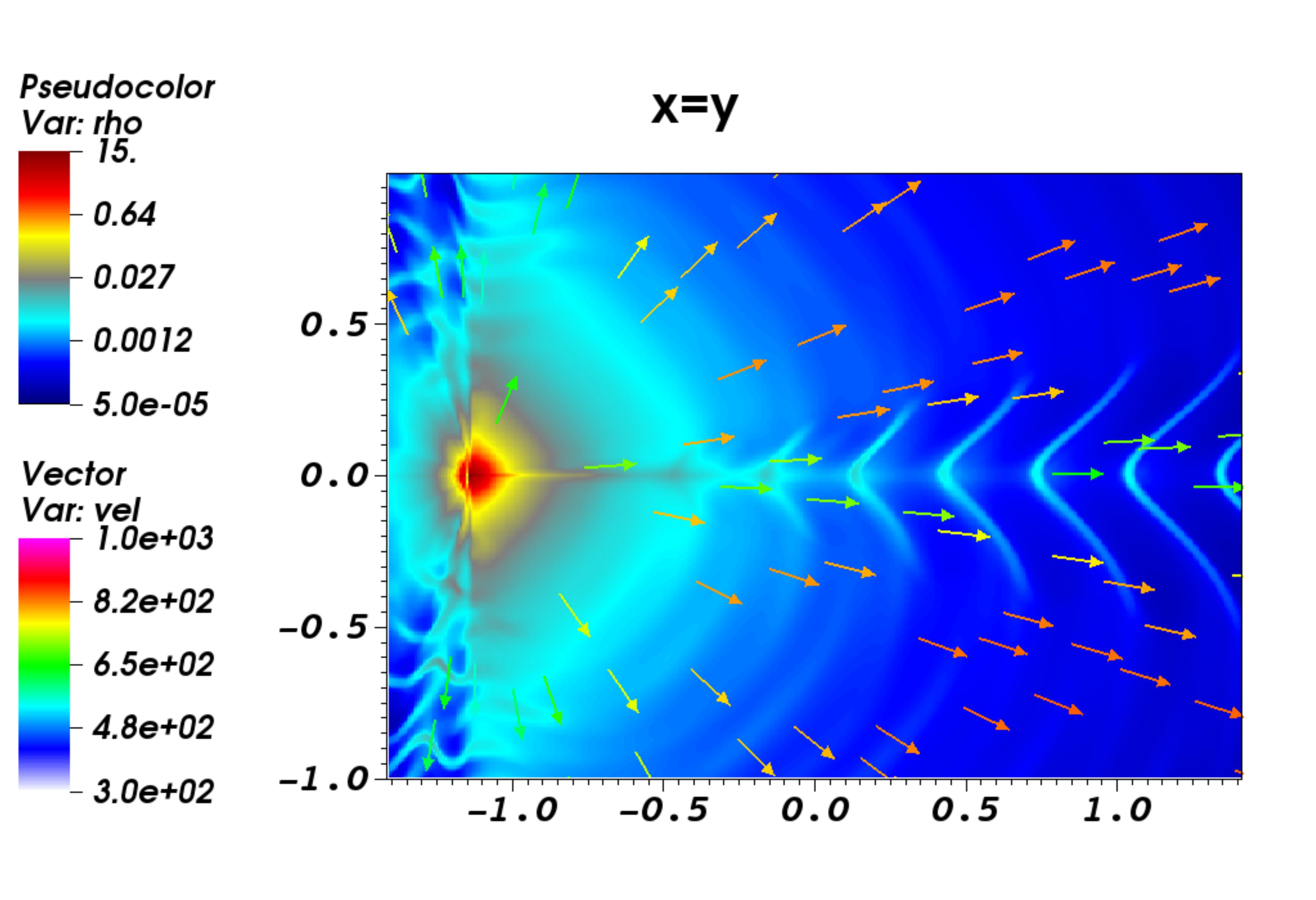}
\caption{Slices of the density and velocity at $z=0$ (top panel) and the plane $x=y$ (bottom panel) 
at time $29.6t_0$ for run {\sf A1}. Unit for density is $\rho_0$ while velocity unit is $1.18\times 10^7$ cm/s. 
The box sizes are in unit of $r_{\rm s}$.}
\label{A1rhoslices}
\end{center}
\end{figure}

A snapshot of 3D density $\rho$ and radiation temperature $T_{\rm r}$
after the collision at time $29.6t_0$ of run {\sf A1} is shown in Figure 
\ref{A1RhoTr}. Here $T_{\rm r}\equiv\left(E_r/a_r\right)^{1/4}$ is the effective blackbody 
temperature of the local radiation field. Due to symmetry, we split the domain in two to simultaneously show $\rho$ and $T_{\rm r}$. The gas redistributes in all directions after the shock as described in 
\cite{Loeb:1997a}. The nearly-equal outgoing mass flow rates through all faces as listed in 
Table \ref{Table:massflux} confirms the roughly symmetric structure.
To see the shock clearly, slices of density and flow velocity through the 
planes $z=0$ and $x=y$ are shown in Figure~\ref{A1rhoslices}. 
The high density stream near $x=y=-0.8r_{\rm s}$ is the location of the shock and most 
of the gas flows within $\sim 80^{\circ}$ along the direction of the net velocity. Notice that a small fraction 
of gas is also injected in the opposite direction of the net velocity because the strong downstream radiation 
pressure accelerates gas in all directions. This is similar to the calculations done by \cite{Kimetal1999}, although 
radiation acceleration was not included appropriately there, as they treated radiation as an effective 
isotropic pressure with a prescribed cooling function.

A high density outflow is ejected perpendicular to the original orbital plane of the stream as shown 
on the left side of Figure~\ref{A1RhoTr}, because the initial obscuration in the perpendicular direction 
is smaller and the radiation acceleration due to the large downstream radiation pressure is larger. 
And despite the fact that the net mass flow rate has no perpendicular component, the flow direction with the largest velocity is $\sim 10^{\circ}-45^{\circ}$ away from the 
original orbital plane $z=0$. The shock front that develops between the two streams is unsteady, with the front oscillating
about the $x=y$ plane; as a result, high density filaments form in the post-shock region, which are clearly visible in Figure~\ref{A1rhoslices}. The periodicity of the instability appears to be on order the stream crossing time $t_{0}$, resulting in a distance between the filaments that is equal to distance the flow travels within 
one oscillation period. The fluctuation of average energy densities shown in Figure~\ref{EnergyHistory} is also caused by the unsteady shock front. The physical 
reason that causes the shock front to oscillate was not initially clear, and to exclude numerical causes, we have checked that the instability exists as we varied the 
the angle that the streams were injected relative to the orientation of the grid (For example, one stream is aligned with the $x$ axis while the other 
stream is $135^{\circ}$ with respect to the $y$ axis). We then carried out a test 
hydrodynamic simulation of the collision where we set $\gamma=4/3$, and the instability also existed with this softer equation of state (the shock is so optically thick 
that the photons provide an effective pressure with adiabatic index $4/3$, so the presence of the instability for $\gamma = 4/3$ is not altogether surprising).

We speculate that the instability arises for the following reasons. When the two streams collide, the shock that develops between them has an angle that is initially equal and 
opposite relative to the two streams, this angle determines the obliqueness of the shock relative to the flows and thus the efficiency of conversion from kinetic to thermal energy. However, if there is a slight 
overpressure that develops on one side of the shock, the shock ``tips over'' towards the opposite stream; this tends to make the shock less oblique relative to the other stream and thus results in more conversion 
 of kinetic energy to internal energy on the opposite side of the shock. Now over-pressured, the opposite side pushes back on the shock interface, tipping it back over in the opposite direction (making the angle 
 less oblique relative to the first stream), the process then repeats itself {\it ad infinitum} as the shock angle tips back and forth on a timescale comparable to the stream sound-crossing time. 
When the optical depth drops with decreasing mass flow rate, the downstream photons can diffuse across the shock on a time scale smaller than the sound-crossing time, equalizing the pressure between the two sides of the shock. As a result, the overpressure from radiation is reduced, and the instability is significantly suppressed as seen in simulation {\sf A1m3} (see Section \ref{sec:mdot}). 

\begin{figure}[t]
\begin{center}
\includegraphics[width=1.0\columnwidth]{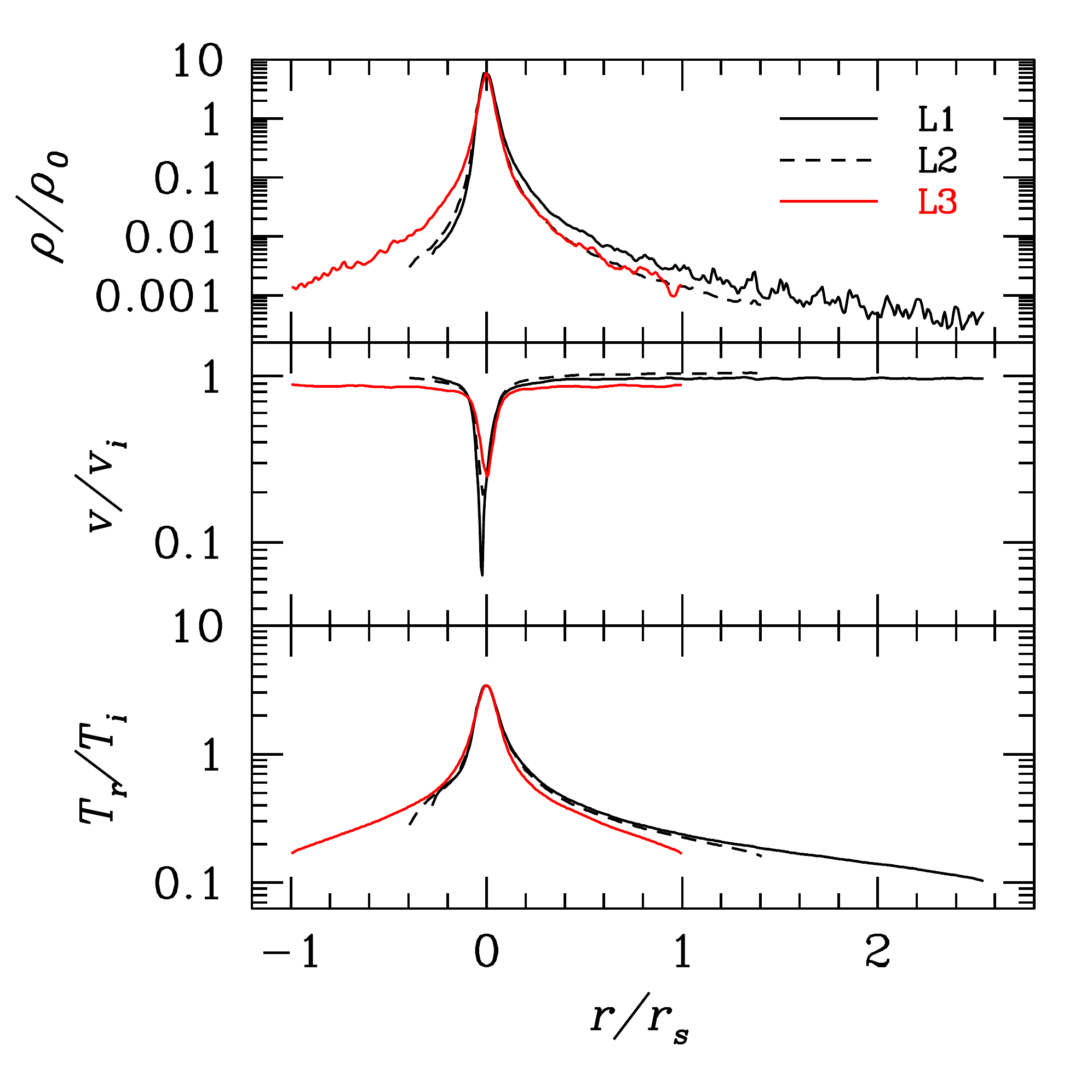}
\caption{Time averaged profiles of density (top panel), flow velocity (middle panel) and radiation temperature (bottom panel) 
through three different lines of sight for simulation {\sf A1}. The time average is done after the initial $10t_0$ ($3.2\times 10^3$ s). The three lines of 
sight are $L1: x=y, z=0$, $L2: x+0.8=y+0.8=z/\sqrt{2}$, and $L3: x=y=-0.8$. The distance to the collision 
point $x=-0.8, y=-0.8, z=0$ is $r$, with the sign represents the direction of flow velocity. All the three lines go through 
the collision point $r=0$. } 
\label{sliceprofile}
\end{center}
\end{figure}

The downstream radiation temperature at the shock is $3.6\times 10^6$ K and it drops to about $10^5$ K near 
the boundary of the simulation box. An order of magnitude estimate of the temperature can be made based on energy conservation, 
where we equate the radiation energy density with the kinetic energy density, $a_rT_s^4\sim \rho_0 v_i^2/2$. Here $T_s$ is the downstream radiation temperature at the shock and 
gas internal energy and upstream thermal energies are 
neglected. For $\rho_0=5.94\times 10^{-9}$ g/cm$^3$ and $v_i=0.31c$, $T_s\sim 4\times 10^6$ K according to this approximation, which is similar to the temperature seen in the post-shock region. Comparing the gas pressure $P_{\rm gas}$ to the radiation pressure $P_{\rm rad}$, we also confirm that $P_{\rm rad} \gg P_{\rm gas}$ at the shock. Energy conservation implies that downstream radiation temperature 
will increase with increasing mass flow rate for a constant $v_i$, but then the effective temperature at the photosphere 
will depend on how the photons are transported to the photosphere, which will be discussed in Section \ref{sec:mdot}.

Time-averaged profiles of the density, flow velocity and radiation temperature along three different lines of sight passing the collision point 
are shown in Figure~\ref{sliceprofile}. The line $L1$ is along the original orbital plane of the stream $x=y,z=0$ while line 
$L3$ is the $z$ axis at the collision point. The line $L2$ is $45^{\circ}$ with respect to the $z$ axis at the $x=y$ plane. 
These lines of sight all probe the properties of the downstream gas. 
The flow velocity reaches the minimum value at the shock position, where the density and radiation pressure peak. 
The maximum downstream density is enhanced by a factor of $6$ compared with the upstream density, 
which is consistent with the very strong radiation pressure 
dominated shock (Figure 13 of \citealt{Jiangetal2012}). The gas gets re-accelerated very quickly by the strong downstream 
radiation pressure and reaches a roughly constant value far away from the photosphere, where most of the thermal energy 
has already been converted back to the kinetic energy.
Density falls as $r^{-2}$ when 
the velocity reaches a constant.  The velocity along $L1$ and $L3$ are always smaller than $v_i$, as most of the mass is 
injected along these directions as shown in Figure~\ref{A1RhoTr}. However, along $L2$, the flow 
velocity can be larger than $v_i$. This will be the source of most of the unbound gas caused by the stream-stream collision, which will be discussed 
in Section \ref{sec:unbound}.

\begin{table*}[t]
\centering
\caption{Mass Flow Rates through All Faces of the Simulation Box}
\begin{tabular}{cc|cccccc|cccccc|c}
\hline
Label	&	$\dot{M}_{0}/\dot{M}_{\rm Edd}$	&	$\dot{M}_{x1}$	&	$\dot{M}_{x2}$	&	$\dot{M}_{y1}$		&	$\dot{M}_{y2}$		& $\dot{M}_{z1}$	& $\dot{M}_{z2}$ 	& $\dot{M}_{x1,u}$	&    $\dot{M}_{x2,u}$ &  $\dot{M}_{y1,u}$ &  $\dot{M}_{y2,u}$ &  $\dot{M}_{z1,u}$ &  $\dot{M}_{z2,u}$ & $\dot{M}_u$	\\
\hline
{\sf A1}		& 0.11 & 16\%	&  15\% &	16\%	&	15\%	&	19\% 	& 19\%	& 	3.3\%	 & 2.2\%	&  3.5\% 	& 2.2\% & 1.7\% & 1.6\%	& $14.5\%$ \\
{\sf A1z}		& 0.11 & 32\%	&  11\% &	32\%	&	11\%	&	7.2\%	& 7.2\%	& 	0.82\%	&  0.0  	& 1.3\% & 0.00 & 1.7\% & 1.7\%  & 	5.5\% \\
{\sf A2}		& 0.11  &4.2\%	&  27\% &	4.2\% &	27\%	&	19\% 	& 19\%	& 	3.7\%        &  3.0\% 	&  3.8\% & 1.2\% & 0.00 & 0.00  & 11.7\% \\
{\sf A1m1}		&  0.22 &   15\%  &  15\% & 15\% &     15\% &      20\% & 20\%     &     5.4\%   	&  1.7\%    &  4.9\%  &  2.6\%  & 0.96\%  & 0.89\% &  16.5\%\\
{\sf A1m2}		&  $4.4\times10^{-2}$ &   15\%  &  14\% & 14\% &     15\% &     21\% & 21\%     &     2.8\%   		&  0.43\%    &  2.8\%  &  0.84\%  & 0.00  & 0.00 & 6.9\% \\
{\sf A1m3}		&  $1.1\times10^{-2}$ &   11\%  &  15\%  & 6.8\% &   15\%   &     26\% & 26\%     &     0.00  		&  0.00    &  0.00  &  0.00  & 0.00  & 0.00 & 0.00 \\
\hline
\end{tabular}
\label{Table:massflux}
\begin{tablenotes}
\item    Note: The second column is the injected mass flow rate carried with the stream scaled with the Eddington accretion rate as defined in Section 
\ref{sec:setup}. The third to fourteenth columns are the outgoing mass flow rates through the left (with subscript 1) and right (with subscript 2) $x,y,z$ surfaces, 
scaled with the total injected mass flow rate. The variables with subscript $u$ represent mass flow rates carried with the unbound gas. The last column is the fraction 
of unbound mass through all six faces.  
\end{tablenotes}
\end{table*}

\begin{table*}[t]
\centering
\caption{Kinetic and Radiative Energy Fluxes Produced by Collisions}
\begin{tabular}{ccccccccccc}
\hline
Label	&	$L_r$	&	$L_{r,0}$	&	$L_{k,x2}$	&	$L_{k,y2}$		&	$L_{k,z1}$		& $L_{k,z2}$	& $L_{r,x2}$ 	& $L_{r,y2}$	&    $L_{r,z1}$ &  $L_{r,z2}$ 	\\
\hline
{\sf A1}		& $3.5\%$	 &    $1.9\%$  &   $22\%$ &   $21\%$  &  $27\%$ & $27\%$ & $0.65\%$ &  $0.67\%$ & $1.1\%$ & $1.1\%$ \\
{\sf A1z}		& $2.4\%$ &    $1.4\%$ &   $10\%$ & 	$10\%$ &  $6.9\%$ & 6.9\% & $0.36\%$ & $0.36\%$ & $0.84\%$ & $0.84\%$	\\
{\sf A2}		&  $2.6\%$&    $1.4\%$  &   $31\%$ &   $30\%$  &  $18\%$ & $18\%$ & $0.63\%$ &  $0.61\%$ & $0.67\%$ & $0.67\%$ \\
{\sf A1m1}		& $3.2\%$&    $1.6\%$  &   $22\%$ &   $22\%$  &  $27\%$ & $27\%$ & $0.61\%$ &  $0.62\%$ & $0.99\%$ & $0.99\%$ \\ 
{\sf A1m2}		&  $4.3\%$&    $2.4\%$  &   $20\%$ &   $22\%$  &  $27\%$ & $27\%$ & $0.82\%$ &  $0.84\%$ & $1.3\%$ & $1.3\%$ \\ 
{\sf A1m3}		&    $7.7\%$&    $5.0\%$  &   $19\%$ &   $19\%$  &  $27\%$ & $27\%$ & $1.7\%$ &  $2.0\%$ & $2.0\%$ & $2.0\%$ \\ 
\hline
\end{tabular}
\label{Table:energyflux}
\begin{tablenotes}
\item    Note: The first column is the total integrated lab frame radiation flux over right $x,\ y$, top and bottom faces, while the second column is the 
total integrated co-moving frame radiation flux over the same four faces. The outgoing kinetic energy fluxes integrated over each face are listed 
in column three to five while the other columns are the outgoing lab frame radiation fluxes integrated over each face. All these luminosities 
are scaled with the total lab frame radiation and kinetic luminosities from the four faces. 
\end{tablenotes}
\end{table*}

\subsection{The Effects of Vertical Offset}

\begin{figure}[t]
\begin{center}
\includegraphics[width=1.0\columnwidth]{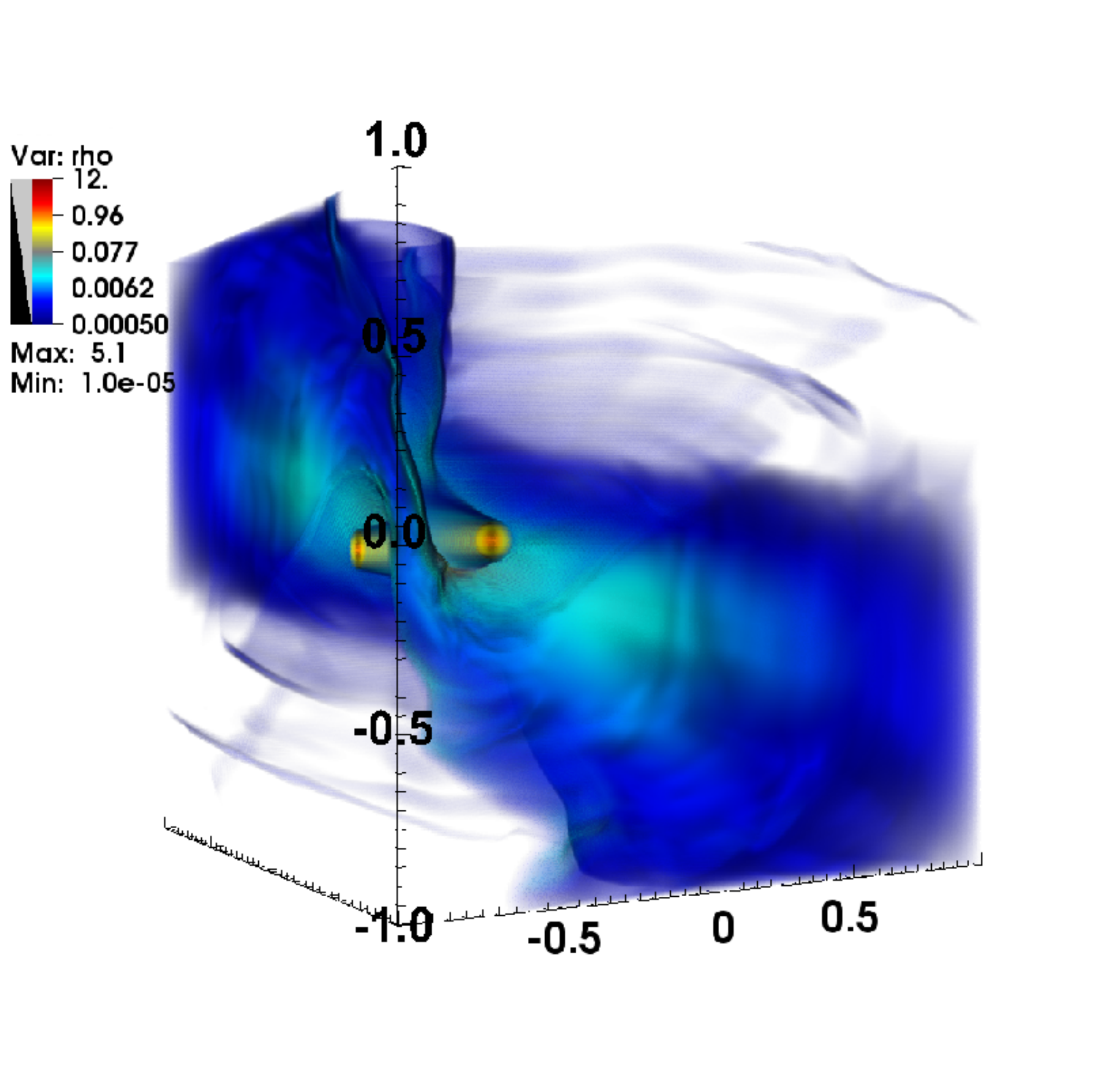}
\caption{Snapshot of density at time $55t_0$ of the simulation {\sf A1z} when a vertical offset is applied. 
Density unit is $\rho_0$ while the box size is in unit of $r_{\rm s}$. (A
movie showing the density evolution for this run is available at \url{ https://youtu.be/jeLUixxE9rA}.)}
\label{3DoffsetRho}
\end{center}
\end{figure}


The black hole spin can cause a vertical offset of the streams when they collide (Section \ref{sec:offset}), which we study 
in simulation {\sf A1z}. The setup is similar to {\sf A1} except that the center of the two streams are 
located at $z=-0.5H$ and $z=0.5H$ respectively. A snapshot of the density at $55t_0$ after the 
collision is shown in Figure~\ref{3DoffsetRho}. Because of the vertical offset and the exponential fall-off of density with distance from the center of the stream, the total momentum flux carried by the stream is 
much larger than the momentum flux of the gas involved in the collision. The primary result of the collision is instead a deflection of the two streams in the vertical direction, and less kinetic energy converted to the thermal energy in the shock 
compared with {\sf A1}. The radiation temperature of the downstream gas at the shock in {\sf A1z} 
is $2.7\times 10^6$ K, $32\%$ of the radiation energy density in {\sf A1}.
 
The vertical offset also causes anisotropic structures at the collision point. The gas that 
does not go through the shock is deflected from the original orbit, which remains cold. 
Table \ref{Table:massflux} confirms that mass flow rates going through the left $x$ and $y$ 
faces (where the streams would have exited the domain in the absence of a collision) are much larger in this case. 
Because of the large optical depth within the stream, photons cannot escape through 
this direction. Instead, most of the radiation is radiated along the direction perpendicular 
to travel direction of the deflected streams. The total optical depth from the photosphere to the collision point is also 
smaller, which is about $600$ along the line of sight $L2$ compared with $2000$ in {\sf A1} 
along the same line of sight. As the downstream radiation energy density and the total optical 
depth are reduced by a similar factor (because they are all roughly proportional the total amount of gas 
involved in the shock), the photosphere temperature along the direction where photons can escape 
will be actually very similar compared with the case {\sf A1}. But because of the reduced photosphere 
radius and the solid angle where photons can escape, a region comparable to the original stream sizes, the radiation efficiency is significantly reduced.

\subsection{Effects of the Collision Angle}


The effects of the collision angles are studied in the run {\sf A2}, where the streams 
collide at $90^{\circ}$ with everything else held the same as {\sf A1}. Due to momentum 
conservation, the net momentum in this case is larger than {\sf A1}, which also means the available 
kinetic energy that can be thermalized  by the shock is smaller. 
We find that the downstream radiation energy density at the shock is $62\%$ of the value in {\sf A1}. 
Most of the downstream gas is confined to a narrow fan-like structure within $20^{\circ}$ along the diagonal line of the 
$x-y$ plane, which is the direction of the net momentum.  As shown in Table \ref{Table:massflux}  and 
\ref{Table:energyflux}, the amount of radiation luminosity and unbound gas produced from the collisions 
are reduced compared with the case {\sf A1}. The case of a smaller impact angle is more likely for deeper, more-relativistic encounters (Equation (\ref{eq:thetac})), so given our result of less conversion of the stream's kinetic energy into outflows and radiation, relativistic encounters are more likely to retain their mass. 

\subsection{Effects of the Injected Mass Flow Rate}
\label{sec:mdot}
\begin{figure}[t]
\begin{center}
\includegraphics[width=1.0\columnwidth]{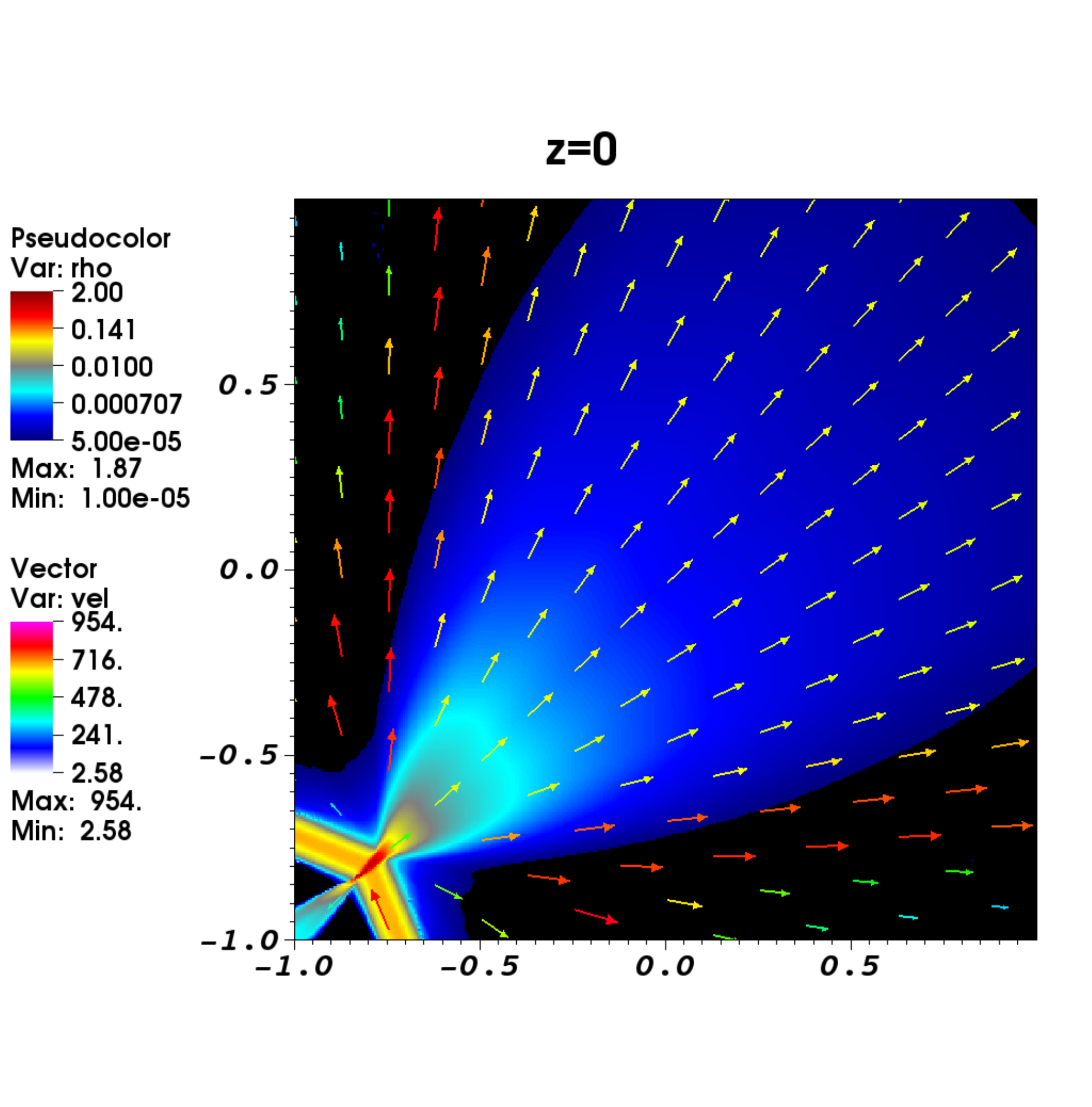}
\caption{Slice of density and flow velocity in the plane of $z=0$ at time 
$63.2t_0$ run {\sf A1m3}. Density and velocity units are $\rho_0$ 
and $1.18\times 10^7$ cm$/$s while the box sizes are in unit of $r_{\rm s}$. 
(A movie showing the density evolution of this run is available at \url{https://youtu.be/NYDUPy-0-2o}.)
} 
\label{3DAng4_rho}
\end{center}
\end{figure}

The four runs {\sf A1m1}, {\sf A1}, {\sf A1m2} and {\sf A1m3} cover
mass flow rates from $22\%$ to $1.1\%$ $\dot{M}_{\rm Edd}$, 
this progression could be seen to be representative of the decline phase of a TDE which is the best-observed period given its prolonged duration \citep[5 years after disruption in the case for][e.g.]{Gezarietal2015}. 
The geometry of the downstream gas is very similar for the four cases as they have 
similar collision parameters. But when the mass flow rate drops to $1.1\%$ $\dot{M}_{\rm Edd}$, 
the shock front becomes 
steady, with the instability we observed so prominently in {\sf A1} disappearing. A slice of density and flow velocity through the $z=0$ plane for run {\sf A1m3} 
is shown in Figure~\ref{3DAng4_rho}. This structure establishes itself quickly after the initial 
collision, and the oscillation of the shock front as shown in Figure~\ref{A1RhoTr} does not 
appear here.



\begin{figure}[t]
\begin{center}
\includegraphics[width=1.0\columnwidth]{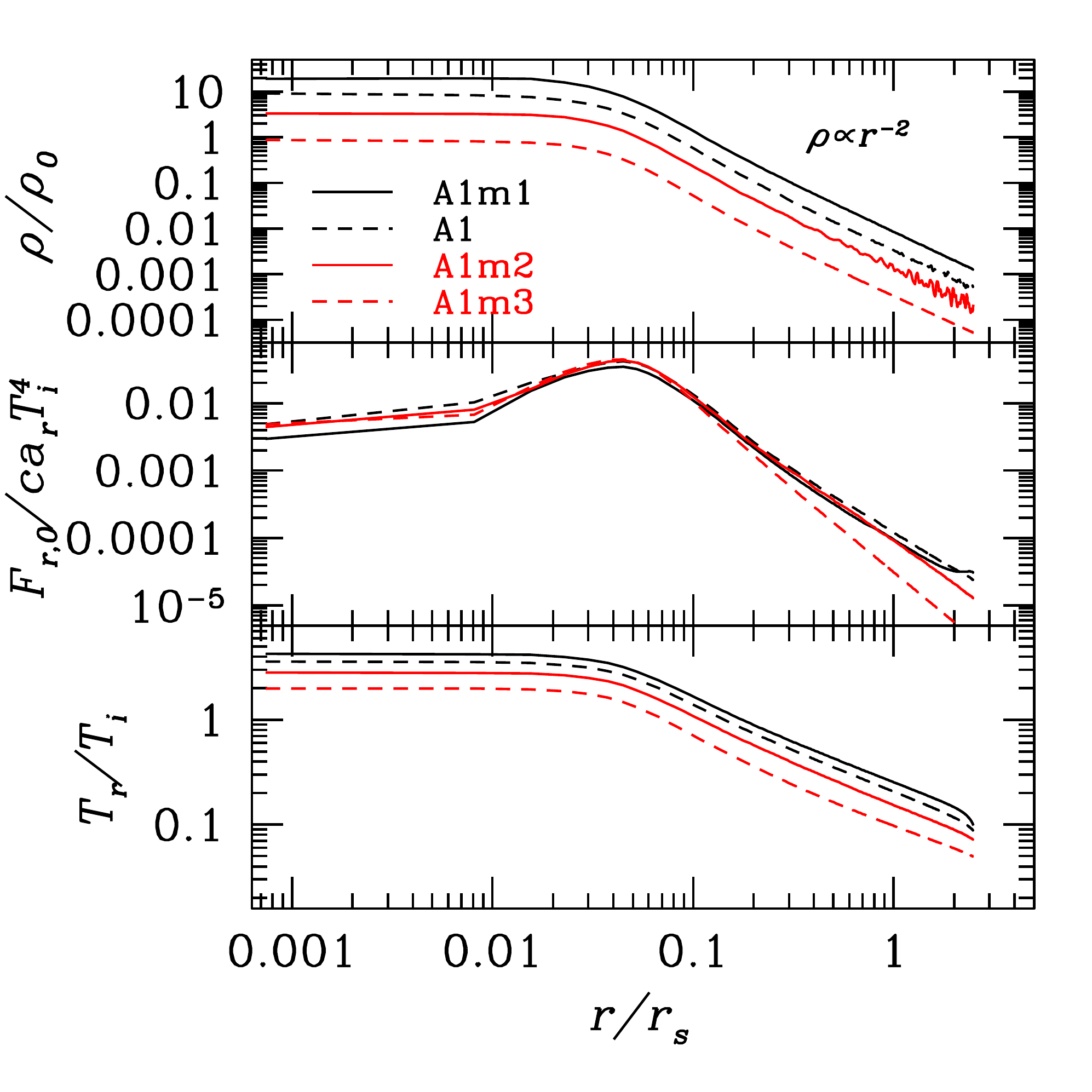}
\caption{Top: radial profiles of the density along the line of sight L1 for runs {\sf A1m1, A1, A1m2, A1m3}. 
The radius $r$ is the distance to the collision point. Middle: radial profiles of the co-moving 
radiation flux along the same line of sight for the four runs. The radiation flux is scaled with $ca_rT_{\rm i}^4$, 
where $T_{\rm i}=10^6$ K. Bottom: radial profiles of the radiation temperature $T_{\rm r}$ for the four runs. } 
\label{compareprofiles}
\end{center}
\end{figure}

We take the line of sight L1 as an example to check the 
radial profiles of density $\rho$, radiation flux in the co-moving frame 
\begin{equation}
F_{r,0}\equiv \sqrt{F_{r,0x}^2+F_{r,0y}^2},
\end{equation}
and radiation temperature 
$T_{\rm r}$ as a function of the distance to the collision point, which
are shown in Figure~\ref{compareprofiles}. 
The density is flat in the acceleration region and it declines with radius roughly as 
$r^{-2}$ when the velocity almost reaches a constant value  (see Figure~\ref{sliceprofile} 
For {\sf A1}).
The density and radiation temperature 
profiles are very similar in the four runs except for their normalization, 
which is proportional to the incoming mass flow rate. The co-moving radiation flux  
also declines with radius roughly as $r^{-2}$, except for run {\sf A1m3}. 
The radiation energy density $E_r$ declines with radius roughly as $r^{-3}$.

The downstream radiation temperatures at the shock $T_{\rm s}$ for the four runs are $4.2\times 10^6$ K, 
$3.6\times 10^6$ K, $2.8\times 10^6$ K and $2.0\times 10^6$ K, which are 
consistent with our estimate based on energy conservation in Section \ref{sec:shock}. 
The radiation temperature in the optically thick part is determined by the diffusion equation,
\begin{eqnarray}
\frac{\partial E_r}{\partial r}=-\frac{3\rho\left(\kappa_a+\kappa_s\right)F_{r,0}}{c}. 
\end{eqnarray}
Here we have assumed the 
Eddington tensor to be $1/3\bI$ for simplicity. If $T_{\rm s}$ is much larger than the temperature $T_{\rm ph}$ at radius at the photospheric radius $r_{\rm ph}$,
\begin{equation}
a_{\rm r}T_{\rm s}^4\approx 3a_{\rm r}T_{\rm ph}^4\tau_{\rm ph},
\label{eq:diffusion}
\end{equation}
where $\tau_{\rm ph}$ is the optical depth from the collision point to the photosphere,
\begin{equation}
\tau_{\rm ph} = \int_0^{r_{\rm ph}} \rho\left(\kappa_a+\kappa_{\rm s}\right)\frac{F_{\rm r,0}}{a_{\rm r}T_{\rm ph}^4}dr.
\end{equation}

Because the photosphere is not entirely within the domain in {\sf A1m1, A1} and {\sf A1m2}, we will compare $\tau_{\rm ph}$ integrated a distance $r_{\rm s}$ instead of $r_{\rm ph}$ (which we denote as $\tau_{\rm s}$) from the collision point to compare between simulations. 
The flux weighted optical depth $\tau_s$ and radiation temperature at $r_{\rm s}$ for {\sf A1m1, A1, A1m2, A1m3} are 
$2.8\times 10^4,\ 3.4\times 10^4,\ 4.3\times 10^4,\ 6.63\times 10^4$ and $2.5\times 10^5$ K, $2.0\times 10^5$ K, $1.5\times 10^5$ K, 
$9.6\times 10^4$ K respectively. They all agree with Equation (\ref{eq:diffusion}) and $T_{\rm s}$ in the four runs very well. 
The ratios between $a_{\rm r}T_{\rm r}^4$ at $r_{\rm s}$ for the four runs are very similar to the ratios of $a_{\rm r}T_{\rm s}^4$, which are also 
the ratios of the incoming mass flow rate. 

When the density follows $\rho_i r_i^2/r^2$, where $r_i$ is a selected radius with density $\rho_i$, 
 the total optical depth integrated from infinity to any radius $r$ for electron scattering opacity is 
$\tau_{\rm es}=\rho_ir_i^2\kappa_{\rm es}/r$. The photosphere radius can be estimated as,\footnote{We take the radius where election scattering optical depth $\tau_{\rm es}$ reaching $1$ 
as the location of photosphere. } 
$r_{\rm ph}\approx \rho_i r_i^2\kappa_{\rm es} $, which is $47r_{\rm s}$, $24r_{\rm s}$, $7r_{\rm s}$ and $1.9r_{\rm s}$ 
for the four runs along line of sight L1 respectively. Assuming that $E_r$ follows the same scaling relation as the photospheric radius, 
the photosphere temperature $T_{\rm ph}$ along the line of sight L1 for {\sf A1m1, A1, A1m2} is estimated to be $1.4\times 10^4$~K, 
$1.8\times 10^4$~K and $3.5\times 10^4$~K. For {\sf A1m3}, the photosphere temperature is $6.2\times 10^4$~K. A similar analysis along line of sight L2 shows that the photosphere radii are $14r_{\rm s}$, $7.0r_{\rm s}$, $2.9r_{\rm s}$ and $1.3r_{\rm s}$, while 
the photosphere temperatures are $3.3\times 10^4$~K, $4.5\times 10^4$~K, $6.3\times 10^4$~K, and $7.2\times 10^4$~K for the four runs. 
Notice that $r_s=2.6\times 10^{12}$ cm for $10^7\msun$ black hole. 

\begin{figure}[t]
\begin{center}
\includegraphics[width=1.0\columnwidth]{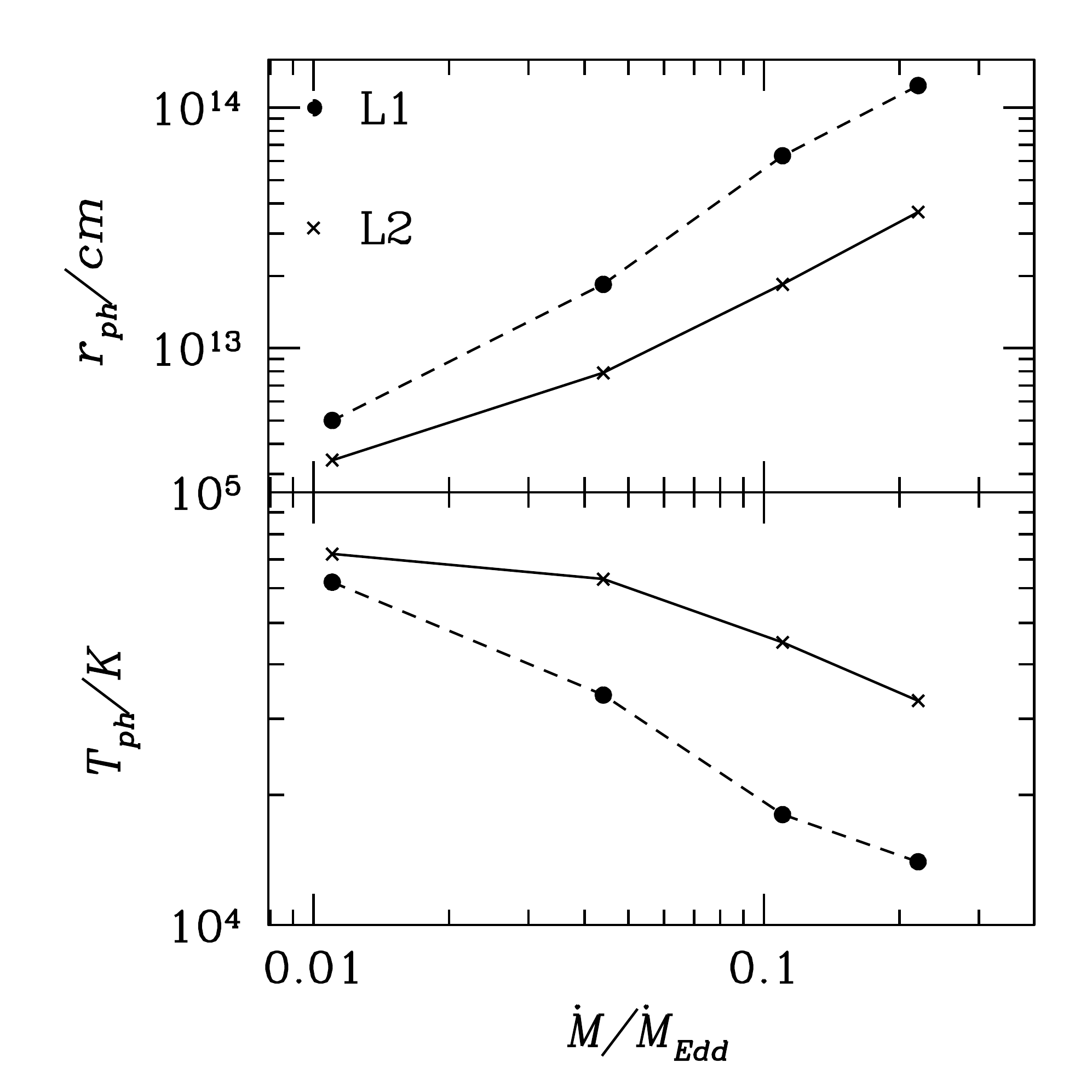}
\caption{Electron scattering photosphere radius $r_{\rm ph}$ (top panel) 
and photosphere temperature (bottom panel) along lines of slight L1 and L2 
as a function of injected mass flow rates for runs {\sf A1m1, A1, A1m2, A1m3}. } 
\label{rTphmdot}
\end{center}
\end{figure}

The electron scattering photosphere radius and photosphere temperature as a function of mass flow rate for the four runs are summarized in Figure \ref{rTphmdot}. For L1, a fit to the photospheric radius and temperature as functions of $\dot{M}$ shows that they are roughly proportional 
to the incoming mass flow rate as
\begin{align}
r_{\rm ph, L1} \approx 6.5\times 10^{14}\left(\frac{\dot{M}_0}{\dot{M}_{\rm Edd}}\right)^{1.09}~{\rm cm}\label{eq:rL1}\\
T_{\rm ph, L1} \approx 6.2\times 10^3\left(\frac{\dot{M}_0}{\dot{M}_{\rm Edd}}\right)^{-0.51}~{\rm K}\label{eq:TL1}.
\end{align}
The above equations confirm that we only capture the photosphere for run {\sf A1m3}. The scalings between $r_{\rm ph}$ and $T_{\rm ph}$ as a function of the mass flow rate along L2 when $\dot{M}_0>0.04\dot{M}_{\rm Edd}$ (we ignored the first point in our fit as the geometric effects change the relationship for lower $\dot{M}$) are 
\begin{align}
r_{\rm ph, L2} \approx 1.6\times 10^{14}\left(\frac{\dot{M}_0}{\dot{M}_{\rm Edd}}\right)^{0.96}~{\rm cm}\label{eq:rL2}\\
T_{\rm ph, L2} \approx 1.8\times 10^4\left(\frac{\dot{M}_0}{\dot{M}_{\rm Edd}}\right)^{-0.4}~{\rm K}\label{eq:TL2}.
\end{align}

Notice that $a_{\rm r}T_{\rm ph}^4 r_{\rm ph}^2$ from line of sight L2 is more than four times larger than the value along line of sight L1, which means more flux is emitted from the hotter surface nearer to the collision point, despite the reduction in surface area. Although the temperature will depend on the exact line of sight to the observer, the larger flux from the hotter line of sight suggests that a typical $T_{\rm ph}$ will tend closer to the values indicated by the expression for L2 (Equation (\ref{eq:TL2})) than the expression for L1 (Equation (\ref{eq:TL1})). For a given mass flow rate $\dot{M}$,  as $a_rT_s^4\propto \dot{M}^2/v_i$ and photosphere radius $\propto \dot{M}/v_i$, 
if we assume the same scaling relation for $E_r$, the photosphere temperature will change with stream velocity as 
$\sqrt{v_i}$. We note however that the observed photosphere temperature of TDE candidates should correspond to the radiation 
temperature at the photosphere of effective absorption opacity, which will be discussed in Section \ref{sec:discussion}.

The total radiation and kinetic energy fluxes leaving from different faces of the simulation box are summarized in 
Table \ref{Table:energyflux}.  The radiation luminosity from the top and bottom surfaces of the simulation box is larger 
than the radiation luminosity from the $x$ and $y$ faces except for run {\sf A1m3}, 
as most of the mass is injected along the orbital plane of the 
original stream. This is consistent with the differences between lines of sight L1 and L2, 
which show that the photosphere temperature from the top and bottom surfaces is larger than 
the photosphere temperature from the $x,y$ faces.

\subsection{The Unbound Gas}
\label{sec:unbound}
The initial stream is in the parabolic orbit around the black hole, which means the kinetic energy 
is comparable to the gravitational potential energy. During the stream-stream collision, the kinetic energy 
is redistributed among the downstream gas via the shock. If part of the gas 
receives more specific kinetic energy than the others as shown in Figure~\ref{sliceprofile}, this part becomes 
unbound to the black hole while the others become more tightly bound. Because the black hole is not included 
in our calculations, we cannot evaluate the final fate of the bound and unbound gas (see Section \ref{sec:discussion}). However, as kinetic energy is much larger 
than the thermal energy after the re-acceleration stops for the downstream gas, we can use the velocity 
alone to give a rough estimate of the fraction of unbound gas. 
If the flow velocity when the gas leaves the simulation box is larger than $v_i$, which by definition is at near-zero binding energy initially, we label it to be unbound. 

The mass flow rates carried with the unbound gas are summarized in Table \ref{Table:massflux}. 
The fractions of the unbound gas for the four runs {\sf A1m1, A1, A1m2} and {\sf A1m3} are 
$16\%$, $14\%$, $6.9\%$ and $0$ respectively. If all the unbound gas escapes to infinity, the kinetic 
energy luminosities carried by the unbound gas are $0.32\%$, $0.46\%$ and $9.8\times 10^{-5}$ of the 
total injected kinetic energy luminosity for runs {\sf A1m1, A1} and {\sf A1m2}. 
This shows that the unbound gas will only form when 
the mass flow rate is larger than $\sim 10\%\dot{M}_{\rm Edd}$ and it increases with $\dot{M}$ above this threshold. 
If we only consider the gas leaving the top and bottom boundaries, which is unlikely to interact with the bound gas which is mostly confined to the original orbital plane, the fractions are reduced to $2\%$ and $4\%$ for {\sf A1m1} and {\sf A1}. 
As expected, we find that there is significantly more unbound gas produced 
with collision angle $\theta=135^{\circ}$ than the case of $90^{\circ}$. For the case with a vertical offset ({\sf A1z}), less of the material is unbound than the zero offset case for the simple reason that a smaller fraction of the gas is involved in the collision, although a larger fraction of the gas is lost in directions perpendicular to the original orbital plane.

In all cases, the stream-stream collision clearly redistributes angular momentum within the two streams \citep{Ramirez-Ruiz:2009a}. The downstream gas is redistributed in basically all directions, with only a small fraction of the gas remaining within the original orbital plane. However, without including the black hole in the simulation box, we cannot calculate the angular momentum exchange self-consistently. This will be the focus of future studies. 

\begin{figure*}[t]
\begin{center}
\begin{tabular}{cc}
{\large \bf Side view} & \includegraphics[align=c,width=0.7\linewidth]{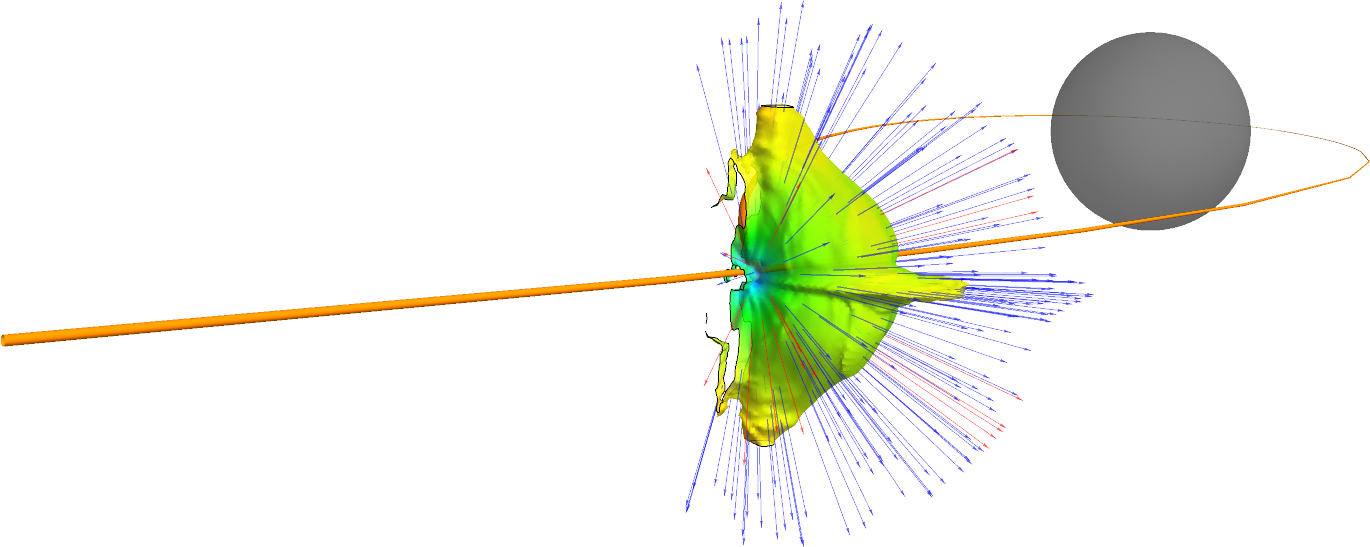}\\
{\large \bf Top view} & \includegraphics[align=c,width=0.7\linewidth]{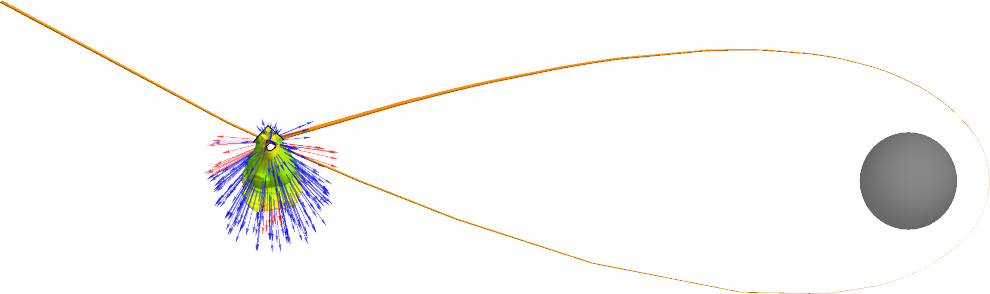}
\end{tabular}
\caption{Three-dimensional rendering of simulation {\sf A1m1} where the simulation domain has been super-imposed onto the global context which determined the stream collision conditions (as shown in Figure~\ref{setup}). The density isosurface shown corresponds to $\rho = 10^{-3} \rho_{0}$, with the color-coding of the surface being proportional to the radiation temperature, ranging from blue (cold) to yellow (hot). The arrows emanating from the isosurface indication the velocity vectors of the portions of the isosurface they originate from, with blue arrows corresponding to material that is bound to the SMBH, and red corresponding to unbound material. It is clear that while the unbound debris may have the velocity required to escape the black hole, its path may either be impeded by bound debris, or that its trajectory may lead to direct accretion onto the black hole.} 
\label{global}
\end{center}
\end{figure*}

\begin{figure*}[t]
\centering
\begin{tabular}{|c|c|c|}
{\large $\dot{M} = 0.22 M_{\rm Edd}$} & {\large $\dot{M} = 0.5 M_{\rm Edd}$} & {\large $\dot{M} = M_{\rm Edd}$}\\
\includegraphics[align=c,width=0.25\linewidth]{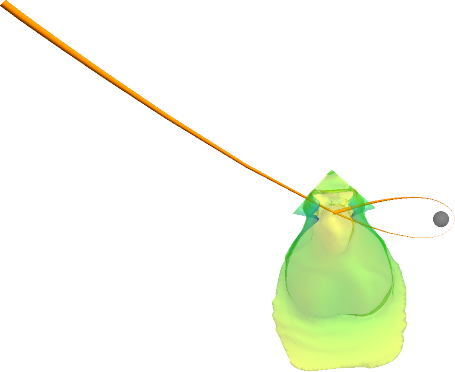} &
\includegraphics[align=c,width=0.25\linewidth]{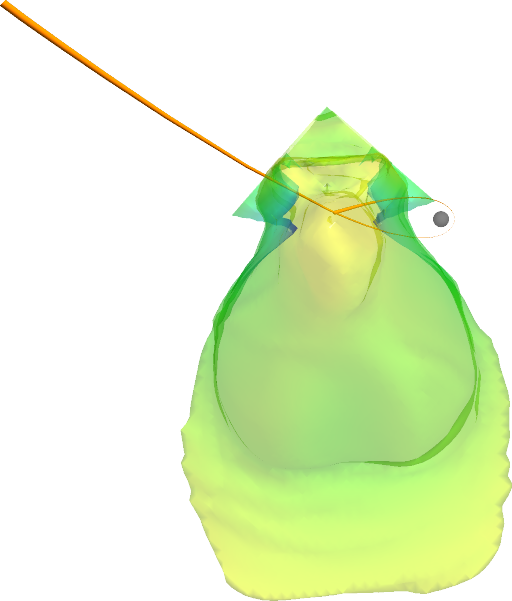} &
\includegraphics[align=c,width=0.25\linewidth]{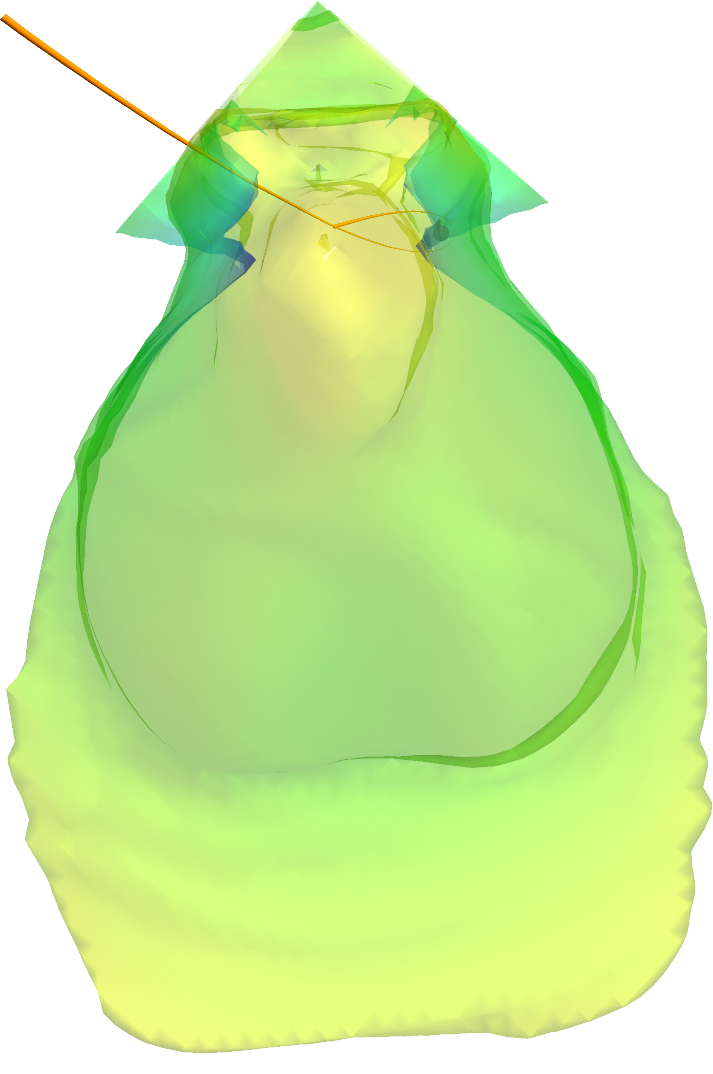}
\end{tabular}
\caption{Photosphere geometry relative to colliding streams for $\dot{M} = 0.22 M_{\rm Edd}$ (corresponding to {\sf A1m1}), $\dot{M} = 0.5 M_{\rm Edd}$, and $\dot{M} = M_{\rm Edd}$. In this figure we have taken the density contour evaluated at $\rho = 10^{-3} \rho_{0}$ (as shown in Figure \ref{global}) and scaled its size to be equal to the photosphere size, using Equations (\ref{eq:rL1}) and (\ref{eq:rL2}), resulting in $r_{L1,ph} = 1.2 \times 10^{14}$, $3.1 \times 10^{14}$, and $6.5 \times 10^{14}$~cm in the above panels (left to right). The stream diameter has been exaggerated by a factor of 5 in this figure to make it visible. At large accretion rates, the outflow from the stream collision point envelopes the black hole, although we stress that the gravity of the black hole needs to be taken into account to determine the true global geometry of the photosphere.} 
\label{accrateglobal}
\end{figure*}

\subsection{Radiation Efficiency}
\label{sec:radeff}
Although most of the radiation energy is quickly converted back to kinetic energy near the shock, a small 
fraction remains as radiation and the photons can escape the photosphere and contribute to the observed luminosity of a flare. The radiation produced in this way is 
directly connected to the initial kinetic energy of the stream without any time delay. Therefore, if the rate of mass fallback onto the black hole follows a given function, the stream-stream collision can produce similar time dependence 
for the emitted radiation field, especially at early times when little matter has accumulated about the black hole from previously circularized material. At late times this could also reproduce the canonical $t^{-5/3}$ decay behavior expected for a large fraction of TDEs, but this may be subject to the additional complication that the black hole may have acquired a significant bound component \citep[\'{a} la][]{Loeb:1997a} that may interfere with the stream collision process.

To see how efficiently the kinetic energy can be converted to the radiation energy during the stream-stream 
collision, we first calculate the total outgoing kinetic and radiation energy fluxes integrated over the right $x$, right $y$, 
top and bottom faces; we do not include the left $x$ and $y$ faces because they are too close to the collision point. 
We define the efficiency as the ratio between the radiation flux and the total energy flux through the four faces, which 
are summarized in Table \ref{Table:energyflux}. 
For runs {\sf A1m1}, {\sf A1}, {\sf A1m2}, {\sf A1m3}, the ratio between the total luminosity 
carried by the lab frame radiation fluxes and the total energy fluxes is $3.2\%$, $3.5\%$, $4.3\%$ and $7.7\%$.  
For runs we do not capture the photosphere, this is just a upper limit for radiation efficiency, in principle a further fraction of the radiation could be converted back to kinetic energy before the photons escape to infinity. 
The lab frame radiation flux is the sum of the co-moving frame radiation flux and the advective radiation enthalpy, 
which can be converted to the kinetic energy flux within the photosphere. To estimate the lower limit of the radiation 
efficiency, we also calculate the ratios between the total luminosity carried by the diffusive radiation fluxes and 
the total energy fluxes, which is $1.6\%$, $1.9\%$, $2.4\%$ and $5.0\%$ for the four runs. This shows that the 
actual radiation efficiency varies between $\sim 2-7\%$ and it increases weakly with decreasing mass flow rate.

\section{Discussions and Conclusions}
\label{sec:discussion}
The simulations we have performed are centered about the stream collision point, but as shown in Figure~\ref{setup}, the domain of the simulation may constitute a small fraction of the volume enclosed within the radius defined by the distance between the black hole and the collision. Gravity has been neglected in this work as we focused on the structure near the shock. As shown in Figure~\ref{sliceprofile}, the distance 
from the shock within which most of the radiation energy is converted back to the kinetic energy is much smaller than $r_{\rm c}$, and the inclusion of gravity will unlikely play any role to change this process. But the gravity can influence the dynamics of the gas once it leaves the simulation domain; an example of a non-trivial interaction that can take place between the bound and unbound debris is shown in Figure~\ref{global}, which was generated from our simulation {\sf A1m1} for which roughly $16\%$ of the post-collision mass was estimated to be unbound. It is clear from inspecting the top view shown in Figure~\ref{global} that some of the bound debris, which will fall back toward the black hole eventually, lies on a track exterior to the path traveled by the unbound debris, and could impede the unbound debris' escape, with the subsequent interaction resulting in a lower unbound fraction than what we have calculated. Even within the unbound debris, internal interactions between fluid elements of different velocities could result in additional energy dissipation within the outflow, as described in \citet{StrubbeQuataert2009} and \citet{Metzger:2015a}. Additionally, some of the material that is considered to be unbound has a trajectory that likely takes it close to the black hole's event horizon, possibly leading to direct accretion with little radiation release \citep{Svirski:2015a}. These effects may lead to a lower unbound fraction than what we have calculated in Section \ref{sec:unbound}.

As mentioned, our simulation domain is too small to capture the photosphere for our higher-$\dot{M}$ simulations. But as shown in Figure~\ref{compareprofiles}, our density profiles fall as $r^{-2}$ from the collision point, suggesting that we resolve the outflow to the point at which the flow becomes homologous; this means that density isocontours should be self-similar beyond our domain (aside from the effects of the black hole's gravity). And as our highest-$\dot{M}$ simulation {\sf A1m1} shows a nearly isotropic ejection of matter, we do not expect the geometry of the outflow to change appreciably for accretion rates larger than what we have presented here. Under the assumption of homology, Figure~\ref{accrateglobal} shows that at higher accretion rates that the photosphere of the outflow produced from the stream collision point can completely envelope the black hole when $\dot{M} \sim \dot{M}_{\rm Edd}$. Once this global structure develops, the energy generated at the point of collision will be reprocessed along with radiation produced by the developing accretion disk. In these instances, the photosphere size may be sufficiently large to match the observed photosphere sizes of events like PS1-10jh, and additionally match the observed rapid evolution of the photosphere size accompanied with increases and decreases in the accretion rate \citep[see Figure 6 of][]{Guillochon:2014a}.

As an example, we consider the conditions that would be required in order to produce the $\gtrsim 2\times 10^{44}$ erg/s observed at peak for PS1-10jh (corresponding to 15\%~$L_{\rm Edd}$ for a $10^7\msun$ black hole) \citep[][]{Gezarietal2012}. Given the 2\% efficiency of converting kinetic energy to radiation that we find at the highest accretion rates, colliding streams with velocity $\sim 0.3c$ would need $\dot{M} \gtrsim 8\dot{M}_{\rm Edd}$ in order to produce the observed luminosity. 
We also note that we adopted the electron scattering and free-free opacity here for simplicity, but with temperatures $\gtrsim$ several $10^4$ K and density 
$\sim 10^{-8}-10^{-9}$ g cm$^{-3}$, the additional opacity provided by the bound-bound transitions of iron-group elements can be much larger than the electron 
scattering value \citep[][]{Jiangetal2015}. This additional opacity may be further supplemented by the bound-free transitions of hydrogen and helium \citep{Roth:2015a}. 
While the inclusion of these additional sources of opacity are unlikely to change the properties of the very optically thick region where the gas and photons 
were already tightly coupled, it will increase the total optical depth. As the observed photosphere should correspond to the location where the total optical depth for effective absorption 
opacity $\sqrt{\kappa_s\kappa_a}$ is one, when absorption opacity is larger than the electron scattering opacity, it will decrease the photosphere temperature relative to what is predicted 
by the simulations we present here. This could result in a larger photosphere that may potentially rival the distance to the black hole, which could potentially explain the large photosphere sizes observed for most optically-selected TDEs. 

Based on the 3D radiation hydrodynamic simulations we have presented in this work, we show that the shock completely changes the structures of the incoming 
stream and redistributes the kinetic energy among the downstream gas. When the mass flow rate 
of the stream is larger than 10\%~$\dot{M}_{\rm Edd}$, this can cause a significant fraction of the 
incoming mass to have specific kinetic energy larger than the initial value and thus becomes unbound to the black hole, 
while the remaining gas becomes more tightly bound. This process also converts $2-8\%$ of the initial kinetic 
energy to the radiation energy that eventually leaves the system as the observed flare. The kinetic energy 
luminosity carried by the unbound gas when it reaches infinity is $\lesssim 0.4\%$ of the initial injected 
kinetic energy rate, which corresponds to $\lesssim 5\times 10^{42}$ erg/s when $\dot{M}_0=\dot{M}_{\rm Edd}$ and $v_i=0.3c$ 
for $10^7\msun$ black hole. The photosphere temperature 
from the stream-stream collision is also consistent with the inferred photosphere temperature of optically discovered TDE candidates. 
 As pointed out by \cite{Piranetal2015}, the flares from stream-stream 
collisions can also follow the same time evolution as the mass flow rate, because there is almost no time delay between 
the collision and the productions of the photons. However, the stream collision would have difficulty simultaneously producing X-rays and optical flux; while there is some variation in temperature along different lines of sight, the variation is not extreme enough to simultaneously produce optical and X-ray emission. This suggests for flares where both components have been observed, such as ASAS-SN 14li \citep{Holoien:2015a,Miller:2015b,Alexander:2015a}, that a secondary process must be responsible for the X-ray emission if the stream collision is responsible for the optical/UV flux.

Because of the computational cost of executing hydrodynamical simulations with self-consistent radiation transfer, we chose to use a small box in order to fully resolve the post stream collision shock region. As the photosphere size will be $\sim 200$ times larger than our current simulation domain when the mass flow rate reaches $\dot{M}_{\rm Edd}$, it is too expensive to simultaneously resolve the collision region and the photosphere within the same simulation at a fixed resolution. Additionally, we neglected to include the gravitational effects of the black hole, which will likely influence the overall shape of the photosphere, especially for higher accretion rates where the optically-thick region encloses the SMBH itself. Lastly, after the original stream is destroyed by the collision, it will not come back to the simulation box anymore, at least not with a structure that is identical to the component of the stream that is returning to periapse for the first time. However, because the orbital period around $r_{\rm c}$ is $\sim 87t_0$ where 
$t_0\equiv r_s/v_i=3.2\times 10^2$ s, the steady state structures within the stream collision region are established on a timescale that is much shorter than the time it takes for the stream collision angle to change due to deflection. This suggests that the collision process may be cyclical with a cycle period equal to roughly this period, resulting in times where the plume is being actively produced and other times where the two streams are not directly interacting (this effect can be seen to some degree in the highest-eccentricity simulations of \citealt{Bonnerot:2015a}). This effect is likely important to consider in the early phases of the stream-stream interaction, but evaluating the nature of this longer timescale variability likely requires a global simulation.

In this paper we performed three-dimensional hydrodynamical simulations with self-consistent radiation transfer which showed that a significant fraction of the kinetic energy carried by the debris streams can be released as observable radiation, with a photosphere size $r_{\rm ph}$ and temperature $T_{\rm ph}$ that depends directly on the rate of accretion $\dot{M}$, and typical $r_{\rm ph}$ and $T_{\rm ph}$ values that are similar to observed optical/UV TDEs. As argued above, the full description of the stream collision process and its emergent radiation will likely required a global simulation to resolve, but the simulations we present here should capture most of the salient features associated with stream-stream collisions. While still under active development \citep[see e.g.][]{Zhu:2015b}, the addition of adaptive mesh refinement in Athena++ should enable to perform these global-scale simulations in the near future.

\section*{Acknowledgements}
This work was supported by the computational resources provided by the NASA High-End Computing
(HEC) Program through the NASA Advanced Supercomputing (NAS) Division
at Ames Research Center; the Extreme Science and Engineering Discovery
Environment (XSEDE), which is supported by National Science Foundation
grant number ACI-1053575.  Y.F.J. and J.G. are supported by NASA through Einstein
Postdoctoral Fellowship grant number PF-140109 and PF3-140108. 
This work was also supported in part by NSF grant AST-1312034.

\bibliography{tde,library}

\begin{thebibliography}{59}
\expandafter\ifx\csname natexlab\endcsname\relax\def\natexlab#1{#1}\fi

\bibitem[{{Alexander} {et~al.}(2015){Alexander}, {Berger}, {Guillochon},
  {Zauderer}, \& {Williams}}]{Alexander:2015a}
{Alexander}, K.~D., {Berger}, E., {Guillochon}, J., {Zauderer}, B.~A., \&
  {Williams}, P.~K.~G. 2015, arXiv:1510.01226

\bibitem[{{Arcavi} {et~al.}(2014){Arcavi}, {Gal-Yam}, {Sullivan}, {Pan},
  {Cenko}, {Horesh}, {Ofek}, {De Cia}, {Yan}, {Yang}, {Howell}, {Tal},
  {Kulkarni}, {Tendulkar}, {Tang}, {Xu}, {Sternberg}, {Cohen}, {Bloom},
  {Nugent}, {Kasliwal}, {Perley}, {Quimby}, {Miller}, {Theissen}, \&
  {Laher}}]{Arcavietal2014}
{Arcavi}, I., {Gal-Yam}, A., {Sullivan}, M., {et~al.} 2014, \apj, 793, 38

\bibitem[{Beloborodov(1999)}]{Beloborodov:1999a}
Beloborodov, A.~M. 1999, High Energy Processes in Accreting Black Holes, 161,
  295

\bibitem[{Bloom {et~al.}(2011)Bloom, Giannios, Metzger, Cenko, Perley, Butler,
  Tanvir, Levan, O'Brien, Strubbe, De~Colle, Ramirez-Ruiz, Lee, Nayakshin,
  Quataert, King, Cucchiara, Guillochon, Bower, Fruchter, Morgan, \& van~der
  Horst}]{Bloom:2011a}
Bloom, J.~S., Giannios, D., Metzger, B.~D., {et~al.} 2011, Science, 333, 203

\bibitem[{{Bonnerot} {et~al.}(2015){Bonnerot}, {Rossi}, {Lodato}, \&
  {Price}}]{Bonnerot:2015a}
{Bonnerot}, C., {Rossi}, E.~M., {Lodato}, G., \& {Price}, D.~J. 2015, ArXiv
  e-prints, 1501.04635

\bibitem[{{Bonnerot} {et~al.}(2016){Bonnerot}, {Rossi}, {Lodato}, \&
  {Price}}]{Bonnerotetal2016}
---. 2016, \mnras, 455, 2253

\bibitem[{Carter \& Luminet(1985)}]{Carter:1985a}
Carter, B., \& Luminet, J.~P. 1985, \mnras, 212, 23

\bibitem[{{Coughlin} \& {Begelman}(2014)}]{CoughlinBegelman2014}
{Coughlin}, E.~R., \& {Begelman}, M.~C. 2014, \apj, 781, 82

\bibitem[{{Coughlin} \& {Nixon}(2015)}]{Coughlin:2015a}
{Coughlin}, E.~R., \& {Nixon}, C. 2015, \apjl, 808, L11

\bibitem[{{Coughlin} {et~al.}(2016){Coughlin}, {Nixon}, {Begelman}, {Armitage},
  \& {Price}}]{Coughlin:2016a}
{Coughlin}, E.~R., {Nixon}, C., {Begelman}, M.~C., {Armitage}, P.~J., \&
  {Price}, D.~J. 2016, \mnras, 455, 3612

\bibitem[{{Dai} {et~al.}(2015){Dai}, {McKinney}, \& {Miller}}]{Dai:2015a}
{Dai}, L., {McKinney}, J.~C., \& {Miller}, M.~C. 2015, \apjl, 812, L39

\bibitem[{{Donley} {et~al.}(2002){Donley}, {Brandt}, {Eracleous}, \&
  {Boller}}]{Donleyetal2002}
{Donley}, J.~L., {Brandt}, W.~N., {Eracleous}, M., \& {Boller}, T. 2002, \aj,
  124, 1308

\bibitem[{{Gezari} {et~al.}(2015){Gezari}, {Chornock}, {Lawrence}, {Rest},
  {Jones}, {Berger}, {Challis}, \& {Narayan}}]{Gezarietal2015}
{Gezari}, S., {Chornock}, R., {Lawrence}, A., {et~al.} 2015, \apjl, 815, L5

\bibitem[{{Gezari} {et~al.}(2006){Gezari}, {Martin}, {Milliard}, {Basa},
  {Halpern}, {Forster}, {Friedman}, {Morrissey}, {Neff}, {Schiminovich},
  {Seibert}, {Small}, \& {Wyder}}]{Gezarietal2006}
{Gezari}, S., {Martin}, D.~C., {Milliard}, B., {et~al.} 2006, \apjl, 653, L25

\bibitem[{{Gezari} {et~al.}(2009){Gezari}, {Heckman}, {Cenko}, {Eracleous},
  {Forster}, {Gon{\c c}alves}, {Martin}, {Morrissey}, {Neff}, {Seibert},
  {Schiminovich}, \& {Wyder}}]{Gezarietal2009}
{Gezari}, S., {Heckman}, T., {Cenko}, S.~B., {et~al.} 2009, \apj, 698, 1367

\bibitem[{{Gezari} {et~al.}(2012){Gezari}, {Chornock}, {Rest}, {Huber},
  {Forster}, {Berger}, {Challis}, {Neill}, {Martin}, {Heckman}, {Lawrence},
  {Norman}, {Narayan}, {Foley}, {Marion}, {Scolnic}, {Chomiuk}, {Soderberg},
  {Smith}, {Kirshner}, {Riess}, {Smartt}, {Stubbs}, {Tonry}, {Wood-Vasey},
  {Burgett}, {Chambers}, {Grav}, {Heasley}, {Kaiser}, {Kudritzki}, {Magnier},
  {Morgan}, \& {Price}}]{Gezarietal2012}
{Gezari}, S., {Chornock}, R., {Rest}, A., {et~al.} 2012, \nat, 485, 217

\bibitem[{{Guillochon} {et~al.}(2014){Guillochon}, {Loeb}, {MacLeod}, \&
  {Ramirez-Ruiz}}]{Guillochon:2014b}
{Guillochon}, J., {Loeb}, A., {MacLeod}, M., \& {Ramirez-Ruiz}, E. 2014, \apjl,
  786, L12

\bibitem[{Guillochon {et~al.}(2014)Guillochon, Manukian, \&
  Ramirez-Ruiz}]{Guillochon:2014a}
Guillochon, J., Manukian, H., \& Ramirez-Ruiz, E. 2014, \apj, 783, 23

\bibitem[{{Guillochon} \& {Ramirez-Ruiz}(2013)}]{Guillochonetal2013}
{Guillochon}, J., \& {Ramirez-Ruiz}, E. 2013, \apj, 767, 25

\bibitem[{{Guillochon} \& {Ramirez-Ruiz}(2015)}]{Guillochon:2015b}
---. 2015, \apj, 809, 166

\bibitem[{Guillochon {et~al.}(2009)Guillochon, Ramirez-Ruiz, Rosswog, \&
  Kasen}]{Guillochon:2009a}
Guillochon, J., Ramirez-Ruiz, E., Rosswog, S., \& Kasen, D. 2009, \apj, 705,
  844

\bibitem[{{Hayasaki} {et~al.}(2015){Hayasaki}, {Stone}, \&
  {Loeb}}]{Hayasakietal2015}
{Hayasaki}, K., {Stone}, N.~C., \& {Loeb}, A. 2015, arXiv:1501.05207

\bibitem[{{Holoien} {et~al.}(2015){Holoien}, {Kochanek}, {Prieto}, {Stanek},
  {Dong}, {Shappee}, {Grupe}, {Brown}, {Basu}, {Beacom}, {Bersier},
  {Brimacombe}, {Danilet}, {Falco}, {Guo}, {Jose}, {Herczeg}, {Long},
  {Pojmanski}, {Simonian}, {Szczygiel}, {Thompson}, {Thorstensen}, \&
  {Wozniak}}]{Holoien:2015a}
{Holoien}, T.~W.-S., {Kochanek}, C.~S., {Prieto}, J.~L., {et~al.} 2015,
  arXiv:1507:01598

\bibitem[{{Jiang} {et~al.}(2015){Jiang}, {Cantiello}, {Bildsten}, {Quataert},
  \& {Blaes}}]{Jiangetal2015}
{Jiang}, Y.-F., {Cantiello}, M., {Bildsten}, L., {Quataert}, E., \& {Blaes}, O.
  2015, \apj, 813, 74

\bibitem[{{Jiang} {et~al.}(2016){Jiang}, {Davis}, \& {Stone}}]{Jiangetal2016}
{Jiang}, Y.-F., {Davis}, S., \& {Stone}, J. 2016, arXiv:1601.06836

\bibitem[{{Jiang} {et~al.}(2012){Jiang}, {Stone}, \& {Davis}}]{Jiangetal2012}
{Jiang}, Y.-F., {Stone}, J.~M., \& {Davis}, S.~W. 2012, \apjs, 199, 14

\bibitem[{{Jiang} {et~al.}(2014{\natexlab{a}}){Jiang}, {Stone}, \&
  {Davis}}]{Jiangetal2014c}
---. 2014{\natexlab{a}}, \apj, 796, 106

\bibitem[{{Jiang} {et~al.}(2014{\natexlab{b}}){Jiang}, {Stone}, \&
  {Davis}}]{Jiangetal2014b}
---. 2014{\natexlab{b}}, \apjs, 213, 7

\bibitem[{{Kim} {et~al.}(1999){Kim}, {Park}, \& {Lee}}]{Kimetal1999}
{Kim}, S.~S., {Park}, M.-G., \& {Lee}, H.~M. 1999, \apj, 519, 647

\bibitem[{Kobayashi {et~al.}(2004)Kobayashi, Laguna, Phinney, \&
  M{\'e}sz{\'a}ros}]{Kobayashi:2004a}
Kobayashi, S., Laguna, P., Phinney, E.~S., \& M{\'e}sz{\'a}ros, P. 2004, \apj,
  615, 855

\bibitem[{Kochanek(1994)}]{Kochanek:1994a}
Kochanek, C.~S. 1994, \apj, 422, 508

\bibitem[{{Kochanek}(2015)}]{Kochanek:2015a}
{Kochanek}, C.~S. 2015, arXiv:1512:03065

\bibitem[{{Komossa}(2015)}]{Komossa2015}
{Komossa}, S. 2015, Journal of High Energy Astrophysics, 7, 148

\bibitem[{{Komossa} \& {Bade}(1999)}]{KomossaBade1999}
{Komossa}, S., \& {Bade}, N. 1999, \aap, 343, 775

\bibitem[{{Krolik} {et~al.}(2016){Krolik}, {Piran}, {Svirski}, \&
  {Cheng}}]{Krolik:2016a}
{Krolik}, J., {Piran}, T., {Svirski}, G., \& {Cheng}, R.~M. 2016,
  arXiv:1602:02824

\bibitem[{Kroupa(2001)}]{Kroupa:2001a}
Kroupa, P. 2001, \mnras, 322, 231

\bibitem[{{Lin} {et~al.}(2015){Lin}, {Maksym}, {Irwin}, {Komossa}, {Webb},
  {Godet}, {Barret}, {Grupe}, \& {Gwyn}}]{Lin:2015a}
{Lin}, D., {Maksym}, P.~W., {Irwin}, J.~A., {et~al.} 2015, \apj, 811, 43

\bibitem[{Lodato {et~al.}(2009)Lodato, King, \& Pringle}]{Lodato:2009a}
Lodato, G., King, A.~R., \& Pringle, J.~E. 2009, \mnras, 392, 332

\bibitem[{Lodato \& Rossi(2011)}]{Lodato:2011a}
Lodato, G., \& Rossi, E.~M. 2011, \mnras, 410, 359

\bibitem[{Loeb \& Ulmer(1997)}]{Loeb:1997a}
Loeb, A., \& Ulmer, A. 1997, \apj, 489, 573

\bibitem[{{MacLeod} {et~al.}(2012){MacLeod}, {Guillochon}, \&
  {Ramirez-Ruiz}}]{MacLeod:2012a}
{MacLeod}, M., {Guillochon}, J., \& {Ramirez-Ruiz}, E. 2012, \apj, 757, 134

\bibitem[{{Metzger} \& {Stone}(2015)}]{Metzger:2015a}
{Metzger}, B.~D., \& {Stone}, N.~C. 2015, arXiv:1506:03453

\bibitem[{{Miller} {et~al.}(2015){Miller}, {Kaastra}, {Miller}, {Reynolds},
  {Brown}, {Cenko}, {Drake}, {Gezari}, {Guillochon}, {Gultekin}, {Irwin},
  {Levan}, {Maitra}, {Maksym}, {Mushotzky}, {O'Brien}, {Paerels}, {de Plaa},
  {Ramirez-Ruiz}, {Strohmayer}, \& {Tanvir}}]{Miller:2015b}
{Miller}, J.~M., {Kaastra}, J.~S., {Miller}, M.~C., {et~al.} 2015, \nat, 526,
  542

\bibitem[{{Miller}(2015)}]{Miller2015}
{Miller}, M.~C. 2015, \apj, 805, 83

\bibitem[{{Phinney}(1989)}]{Phinney1989}
{Phinney}, E.~S. 1989, in IAU Symposium, Vol. 136, The Center of the Galaxy,
  ed. M.~{Morris}, 543

\bibitem[{{Piran} {et~al.}(2015){Piran}, {Svirski}, {Krolik}, {Cheng}, \&
  {Shiokawa}}]{Piranetal2015}
{Piran}, T., {Svirski}, G., {Krolik}, J., {Cheng}, R.~M., \& {Shiokawa}, H.
  2015, \apj, 806, 164

\bibitem[{Ramirez-Ruiz \& Rosswog(2009)}]{Ramirez-Ruiz:2009a}
Ramirez-Ruiz, E., \& Rosswog, S. 2009, \apjl, 697, L77

\bibitem[{{Rees}(1988)}]{Rees1988}
{Rees}, M.~J. 1988, \nat, 333, 523

\bibitem[{Rosswog {et~al.}(2009)Rosswog, Ramirez-Ruiz, \& Hix}]{Rosswog:2009a}
Rosswog, S., Ramirez-Ruiz, E., \& Hix, W.~R. 2009, \apj, 695, 404

\bibitem[{{Roth} {et~al.}(2015){Roth}, {Kasen}, {Guillochon}, \&
  {Ramirez-Ruiz}}]{Roth:2015a}
{Roth}, N., {Kasen}, D., {Guillochon}, J., \& {Ramirez-Ruiz}, E. 2015,
  arXiv:1510.08454

\bibitem[{{Sadowski} {et~al.}(2015){Sadowski}, {Tejeda}, {Gafton}, {Rosswog},
  \& {Abarca}}]{Sadowski:2015b}
{Sadowski}, A., {Tejeda}, E., {Gafton}, E., {Rosswog}, S., \& {Abarca}, D.
  2015, arXiv:1512.04865

\bibitem[{{Shiokawa} {et~al.}(2015){Shiokawa}, {Krolik}, {Cheng}, {Piran}, \&
  {Noble}}]{Shiokawa:2015a}
{Shiokawa}, H., {Krolik}, J.~H., {Cheng}, R.~M., {Piran}, T., \& {Noble}, S.~C.
  2015, \apj, 804, 85

\bibitem[{{Stone} {et~al.}(2008){Stone}, {Gardiner}, {Teuben}, {Hawley}, \&
  {Simon}}]{Stoneetal2008}
{Stone}, J.~M., {Gardiner}, T.~A., {Teuben}, P., {Hawley}, J.~F., \& {Simon},
  J.~B. 2008, \apjs, 178, 137

\bibitem[{{Stone} \& {Loeb}(2012)}]{Stone:2012a}
{Stone}, N., \& {Loeb}, A. 2012, Physical Review Letters, 108, 061302

\bibitem[{{Strubbe} \& {Quataert}(2009)}]{StrubbeQuataert2009}
{Strubbe}, L.~E., \& {Quataert}, E. 2009, \mnras, 400, 2070

\bibitem[{{Svirski} {et~al.}(2015){Svirski}, {Piran}, \&
  {Krolik}}]{Svirski:2015a}
{Svirski}, G., {Piran}, T., \& {Krolik}, J. 2015, arXiv:1508.02389

\bibitem[{{van Velzen} {et~al.}(2011){van Velzen}, {Farrar}, {Gezari},
  {Morrell}, {Zaritsky}, {{\"O}stman}, {Smith}, {Gelfand}, \&
  {Drake}}]{VanVelzenetal2011}
{van Velzen}, S., {Farrar}, G.~R., {Gezari}, S., {et~al.} 2011, \apj, 741, 73

\bibitem[{{Vink{\'o}} {et~al.}(2015){Vink{\'o}}, {Yuan}, {Quimby}, {Wheeler},
  {Ramirez-Ruiz}, {Guillochon}, {Chatzopoulos}, {Marion}, \&
  {Akerlof}}]{Vinko:2015a}
{Vink{\'o}}, J., {Yuan}, F., {Quimby}, R.~M., {et~al.} 2015, \apj, 798, 12

\bibitem[{{Zhu} {et~al.}(2015){Zhu}, {Dong}, {Stone}, \& {Rafikov}}]{Zhu:2015b}
{Zhu}, Z., {Dong}, R., {Stone}, J.~M., \& {Rafikov}, R.~R. 2015, \apj, 813, 88

\end{thebibliography}

\end{CJK*}

\end{document}